\documentclass[preprint,onecolumn,nofootinbib]{revtex4}
\pdfoutput=1
\usepackage{CJK}
\usepackage[colorlinks=true,linkcolor=blue,urlcolor=blue,filecolor=black,citecolor=red,pdfstartview=FitV,pdftitle={},pdfsubject={},pdfkeywords={},pdfpagemode=None,bookmarksopen=true]{hyperref}
\usepackage{graphicx}
\usepackage{amsmath}
\usepackage{amsfonts}
\usepackage{amssymb,ulem}
\usepackage{color}%
\usepackage{dcolumn}
\newcommand{\f}{\begin{equation}}
\newcommand{\ff}{\end{equation}}
\newcommand{\fa}{\begin{eqnarray}}
\newcommand{\ffa}{\end{eqnarray}}

\begin{document}
\title{Informational properties for Einstein-Maxwell-Dilaton Gravity}
\author{Guoyang Fu$^{1}$}
\thanks{FuguoyangEDU@163.com}
\author{Peng Liu $^{2}$}
\thanks{phylp@email.jnu.edu.cn}
\author{Huajie Gong $^{1}$}
\thanks{huajiegong@qq.com}
\author{Xiao-Mei Kuang$^{1}$}
\thanks{xmeikuang@yzu.edu.cn}
\author{Jian-Pin Wu $^{1}$}
\thanks{jianpinwu@yzu.edu.cn}
\affiliation{
\small{$^{1}$Center for Gravitation and Cosmology, College of Physical Science and Technology, Yangzhou University, Yangzhou 225009, China \\
$^{2}$ Department of Physics and Siyuan Laboratory, Jinan University, Guangzhou 510632, China\\}
}

\begin{abstract}
We study the information quantities, including the holographic entanglement entropy (HEE), mutual information (MI) and entanglement of purification (EoP), over Gubser-Rocha model.
The remarkable property of this model is the zero entropy density at ground state, in term of which we expect to extract novel, even singular informational properties in zero temperature limit. Surprisedly, we do not observe any singular behavior of entanglement-related physical quantities under the zero temperature limit. Nevertheless, we find a peculiar property from Gubser-Rocha model that in low temperature region, the HEE decreases with the increase of temperature, which is contrary to that in most holographic models. We argue that this novel phenomenon is brought by the singular property of the zero temperature limit, of which the analytical verification is present.
In addition, we also compare the features of the information quantities in Gubser-Rocha model with those in Reissner-Nordstrom Anti-de Sitter (RN-AdS) black hole model. It is shown that the HEE and MI of Gubser-Rocha model are always larger than those of RN-AdS model, while the EoP behaves in an opposite way. Our results indicate that MI and EoP could have different abilities in describing mixed state entanglement.
\end{abstract} \maketitle

\section{Introduction}

Quantum entanglement is playing an increasingly prominent role in modern physics from condensed matter theory to the black hole theory.
In the context of Anti-de Sitter/Conformal Field Theory (AdS/CFT) correspondence  \cite{Maldacena:1997re,Gubser:1998bc,Witten:1998qj,Aharony:1999ti},
quantum entanglement also plays a key role in the investigation on how the bulk spacetime emerges from the entanglement structure \cite{Maldacena:2001kr,VanRaamsdonk:2010pw,VanRaamsdonk:2018zws,Maldacena:2013xja,Takayanagi:2018pml}.
The entanglement entropy (EE) is a measure of quantum entanglement between two subsystems A and B for a given pure state.
In holographic framework, EE has a simple geometric description known as Rangamani-Takayanagi (RT) formula
that EE for a subregion on the dual boundary is proportional to the minimal surface in the bulk geometry, which is dubbed as holographic entanglement entropy (HEE) \cite{Ryu:2006bv,Takayanagi:2012kg,Lewkowycz:2013nqa}.
For covariant cases, RT formula is reformulated into the Hubeny-Rangamani-Takayanagi (HRT) formula \cite{Hubeny:2007xt,Dong:2016hjy}. Their proposal matches very well with the known results from the two-dimensional CFT \cite{Headrick:2007km,Wall:2012uf,Ryu:2006ef}. The success of RT/HRT formula inspires lots of works toward better understanding of this topic \cite{Casini:2011kv,Rahimi:2016bbv,Lokhande:2017jik,Myers:2012ed} and some important applications.
One of the most important applications of HEE is that it can characterize phase transition,
including quantum phase transitions and thermodynamic phase transitions, see for instances \cite{Ling:2015dma,Ling:2016wyr,Ling:2016dck,Pakman:2008ui,Kuang:2014kha,Klebanov:2007ws,Zhang:2016rcm,
Zeng:2016fsb,Guo:2019vni}.

However, EE suffers from UV divergence in general and one has to use a regularization method to remove the divergence.
To overcome this regulator-dependent measure of entanglement, a special linear combinations of EE called mutual information (MI) was proposed, which is a positive definite quantity guaranteed by the subadditivity and free from the UV
divergences \cite{Casini:2004bw,Wolf:2007tdq,Headrick:2010zt}.
In addition, MI removes the thermal entropy contribution \cite{Fischler:2012uv}.
Therefore, MI is a good probe to learn basic properties of any local observable in a quantum system \cite{Groisman:2004,Wolf:2007tdq}. In holographic framework, one can directly calculate MI with the use of RT/HRT formula, which is called holographic mutual information (HMI).

It is well known that HEE is a good measure for a bipartite pure state, but it is not suitable for measuring mixed state entanglement. A novel quantity called entanglement of purification (EoP) $E_p$ is introduced to depict the measure for mixed state entanglement in \cite{Terhal:2002}. However, since the calculation of EoP is to minimize the HEE in the extremely huge parameter space of purification, it is a hard task to calculate $E_p$ in quantum field theory (QFT) \cite{Caputa:2018xuf}. For holographic dual field theory, it is conjectured that $E_p$ is dual to the area of the minimal cross section of the entanglement wedge $E_w$ \cite{Takayanagi:2017knl,Nguyen:2017yqw,Bao:2018gck,Umemoto:2018jpc}, in which it was addressed that most properties of $E_w$ are consistent with those of $E_p$ in QFT.

Most of these works \cite{Takayanagi:2017knl,Nguyen:2017yqw,Bao:2018gck,Umemoto:2018jpc} are implemented in $3$-dimensional AdS spacetime.
For higher dimensional AdS geometry, we usually resort to the numerics. As a first attempt, the authors of \cite{Yang:2018gfq} numerically studied the properties of EoP and its evolution behavior for
thermofield double states dual to the Schwarzschild  black hole. Later, in \cite{Liu:2019qje}, the authors have developed an algorithm to calculate EoP for symmetric and asymmetric configuration in pure $AdS_4$ and $4$ dimensional Reissner-Nordstrom Anti-de Sitter (RN-AdS) black hole backgrounds. The temperature behavior of EoP and some important inequalities of EoP are numerically explored.
Further, the authors in \cite{Huang:2019zph} studied some holographic informational quantities, including HEE, MI and EoP, and they argued that the EoP may be a better entanglement measure of the mixed state than MI. In addition, the connection between holographic EoP and holographic complexity of purification (CoP) was also explored for various models in \cite{Ghodrati:2019hnn}.

In this paper, we shall study the information quantities, including HEE, MI and EoP in Gubser-Rocha model \cite{Gubser:2009qt}.
The black brane solution of Gubser-Rocha model possesses two important and appealing characteristics, i.e.,
zero ground state entropy density and linear specific heat at low temperature,
which are also the characteristics of a Fermi gas.
The study of the probe fermionic spectrum over Gubser-Rocha model also confirm that the system shares the same property of the degenerate Fermi liquid \cite{Wu:2011cy,Li:2011sh}.
Some interesting studies based on Gubser-Rocha model have also been implemented, see \cite{Ling:2013nxa,Ling:2016wyr,Alishahiha:2012ad} and references therein.
In particular, in \cite{Ling:2013nxa}, the ionic lattice background in the framework of Gubser-Rocha model was constructed and the optical conductivity of the dual field theory on the boundary was studied. In addition, some of us constructed a Q-lattice deformed Gubser-Rocha model with vanishing ground entropy density, over which the HEE was explored and it was claimed that the first order derivative of HEE with respect to Q-lattice parameters could characterize the quantum phase transition \cite{Ling:2016wyr}.

In contrast with RN-AdS geometry, which has a non-vanishing ground state entropy density, Gubser-Rocha model provides a novel platform to study the holographic phenomena. Here, we aim to study the universal properties of HEE, MI and EoP over Gubser-rocha model and compare the results from such a model with vanishing ground state entropy density with that from RN-AdS geometry studied in \cite{Liu:2019qje}.

Our paper is organized as what follows. In Section \ref{sec-GR-model}, we present a brief review on Gubser-Rocha model.
And then, we numerically calculate the HEE, MI and EoP and discuss the novel properties of these holographic informational quantities in Section \ref{sec-irq}.
The conclusions and discussions are presented in Section \ref{sec-con}.

\section{Gubser-Rocha model}\label{sec-GR-model}

We start with the following action \cite{Gubser:2009qt,Gubser:2000mm}
\f\label{DilatonAction}
S= {1 \over 2\kappa^2}\int d^{4}x \sqrt{-g} \left[ R - {1 \over 4} e^{\Phi} F^{\mu\nu}F_{\mu\nu} -
{3\over 2} (\partial_\mu\Phi)^2 + {6 \over L^2} \cosh \Phi \right]\,,
\ff
where $L$ is the AdS radius, $\Phi$ is the dilation field and $F_{ab}=\partial_{a}A_{b}-\partial_{b}A_{a}$.
An analytical charged black brane solution to the above action has previously been given in \cite{Gubser:2009qt}
\fa \label{Metric}
&&
ds^{2} = \frac{L^2}{z^2}\left(-f(z)dt^2+\frac{dz^2}{f(z)}+g(z)(dx^2+dy^2)\right)\,,
\\ \label{GaugeAt}
&& A_{t}(z)=L\sqrt{3Q}(1-z)\frac{\sqrt{1+Q}}{1+Q z}\,,
\ffa
where
\fa
&&
f(z)=(1-z)\frac{p(z)}{g(z)},~~~~g(z)=\left(1+Qz\right)^{3/2}\,,
~~~~
\nonumber\\
&&\label{Metricp}
p(z)=1+\left(1+3Q\right)z+\left(1+3Q\left(1+Q\right)\right)z^{2}\,.
\ffa
The coordinate system we take is same as that in \cite{Ling:2013nxa},
which is the coordinate transformation based on the solution presented in \cite{Gubser:2009qt}.
In our current coordinate system, the Hawking temperature can be worked out as
\f \label{T} \hat{T}
=\frac{3\sqrt{1+Q}}{4\pi L }. \ff
The system is determined by the scaling invariant temperature
\f
\label{HawkingT}
T=\frac{\hat{T}}{\mu} =\frac{\sqrt{3}}{4\pi L
\sqrt{Q}}\,,
\ff
where $\mu$ is the chemical potential in the dual boundary field theory
and it is related to the parameter $Q$ as
\f
\mu=\sqrt{3 Q (1+Q)}\,.
\ff
The temperature $T$ is inversely proportional to the parameter $Q$.
When $Q$ tends to zero, it goes up to infinity and the black brane is a Schwarzschild-AdS black brane.
While as $Q$ approaches to infinity, it goes down to zero, which corresponds to an extremal black brane.

It is worthwhile to emphasize that this black brane geometry possesses an appealing characteristic, i.e., the zero ground state entropy,
in contrast to the usual RN-AdS black brane which have finite entropy density even at the zero temperature.
For more discussion on the thermodynamics of this charged black brane, please refer to \cite{Gubser:2009qt}.

In high temperature limit, the geometry of the Gubser-Rocha model and the RN-AdS model are the same. This can be seen by taking the $Q\to 0$ limit for Gubser-Rocha model and $\mu \to 0$ limit for RN-AdS model.
It predicts that all the information-related quantities will be the same.
As long as the quantities related to quantum information are only related to the background geometry, we can conclude that the quantum information behavior of Gubser-Rocha model is consistent with that of RN-AdS model in the high temperature limit.
Therefore, we pay more attention to their finite temperature and extremely low temperature behavior.

\section{The holographic information related quantities}\label{sec-irq}

In this section, we shall study the holographic information related quantities over Gubser-Rocha model.
In order to demonstrate the peculiar properties of the holographic information quantities over Gubser-Rocha model,
we shall also present the corresponding results from RN-AdS black brane for comparison throughout this paper.
For RN-AdS black brane geometry, we refer to \cite{Liu:2019qje}.

\subsection{Holographic entanglement entropy}
EE, as a measure of entanglement, is one of the hot topics in quantum information.
For a pure state system composed of two parts A and B,
its EE is described by Von Neumann entropy $S_A\equiv -Tr(\rho_A \ln{\rho_A})$, where $\rho_A\equiv Tr_B |\psi\rangle \langle \psi|$.
Holographically, the Von Neumann entropy can be depicted through the R-T formula as \cite{Ryu:2006bv}
\fa
S_A = \frac{\texttt{Area}(\gamma_A)}{4 G_N}\,,
\ffa
where $G_N$ is the bulk Newton constant. $\gamma_A$ is the minimal surface which stretches into the bulk and ends at $\partial A$.

\begin{figure}[ht!]
    \centering
    \includegraphics[width=0.5\textwidth]{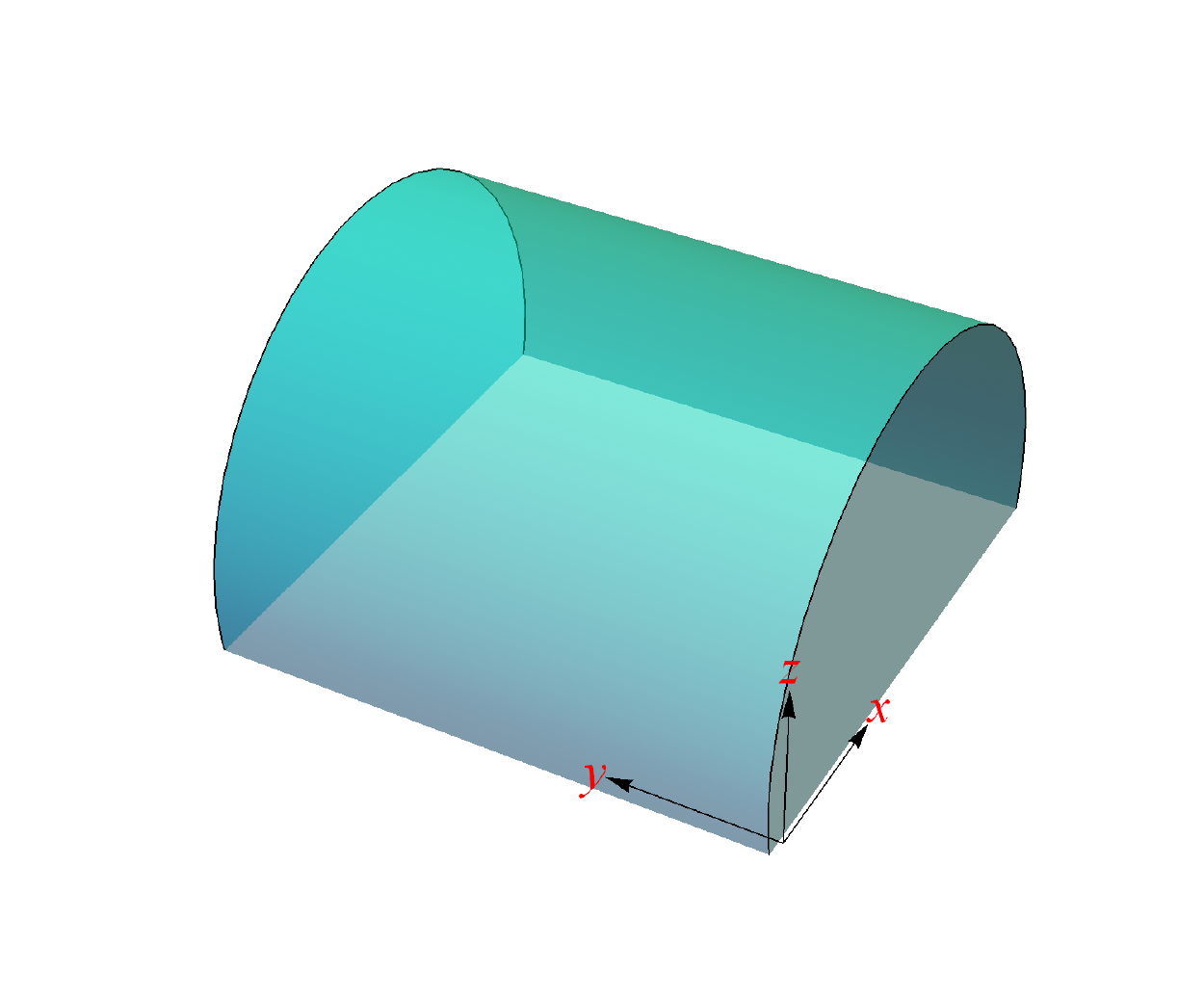}
\caption{Diagram of the extremal surface for an infinite strip configuration with width $l$.}
    \label{EE-cartoon}
\end{figure}
We consider a specific configuration that the subsystem $A$ is an infinite strip along $y$-axis with width $l$ along $x$-axis,
i.e., $A:=\{0 < x < l, -\infty < y < \infty\}$ (see FIG.\ref{EE-cartoon}).
Since the minimum surface is invariant along $y$-axis,
it is convenient to describe this minimum surface by the radial coordinate $z(x)$.
When we fix the width, the HEE can be explicitly expressed as
\fa
&&
\label{S-ori}
\hat{S}=2 \int^{z_*}_{\epsilon} dz \frac{g(z)^{\frac{3}{2}}}{z^2 \sqrt{f(z) (g(z)^2 - \frac{z^4}{z_*^4}g(z_*)^2)}}\,,
\
\\
&&
\label{zs_vs_w}
\hat{l}=2\int_{\epsilon}^{z_*} dz \frac{g(z_*)}{\sqrt{f(z)g(z)}\sqrt{\frac{z_*^4 g(z)^2}{z^4}-g(z_*)^2}}\,,
\ffa
where $z_*$ is the top (alternatively call the turning point) of the minimum surface.
Since HEE is divergent at the asymptotic AdS boundary, we have introduced a cutoff $\epsilon$ in the above expression.
To subtract out the vacuum contribution to HEE, we add a counter term $-1/z^2$ into the integration of $\hat{S}$ such that we have the regularized HEE as what follows
\f\label{hee}
\hat{S}=2 \left( \int^{z_*}_{\epsilon} \Big(\frac{g(z)^{\frac{3}{2}}}{z^2 \sqrt{f(z) (g(z)^2 - \frac{z^4}{z_*^4}g(z_*)^2)}} - \frac{1}{z^2}\Big) dz - \frac{1}{z_*}\right)\,.
\ff
Note that here $\hat{l}$ and $\hat{S}$ are dimensionfull width and HEE.
Adopting $\mu$ as the scaling unit, we have the scaling-invariant width and HEE, which are $l\equiv\hat{l}\mu$ and $S\equiv\hat{S}/\mu$.
We shall only focus on the scaling-invariant quantities in the following study.

\begin{figure}[ht!]
    \centering
    \includegraphics[width=0.45\textwidth]{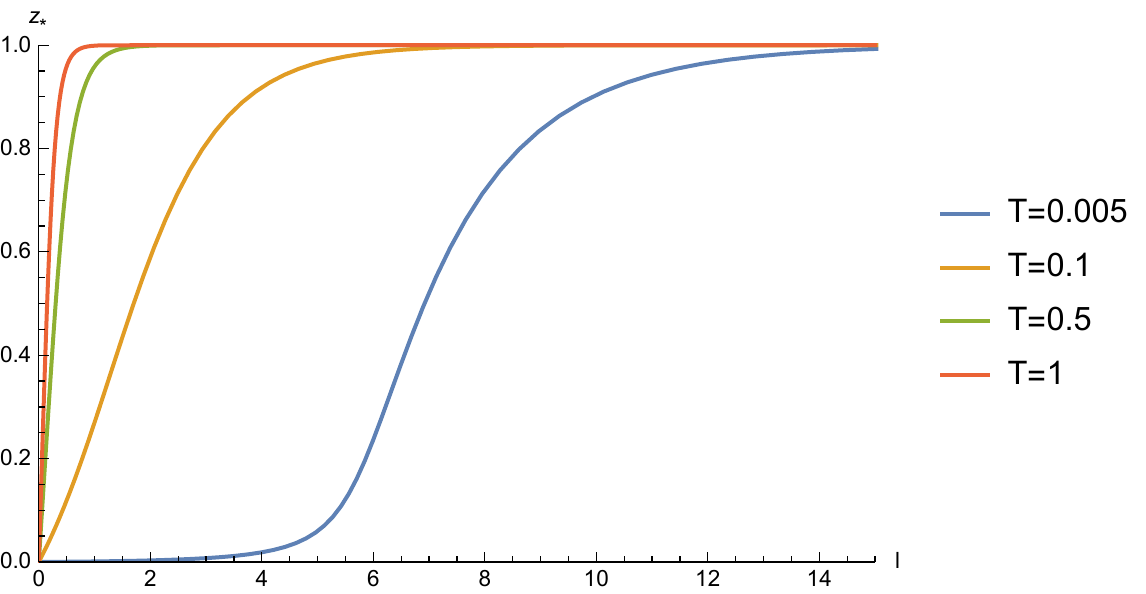}\ \hspace{0.2cm}
    \includegraphics[width=0.45\textwidth]{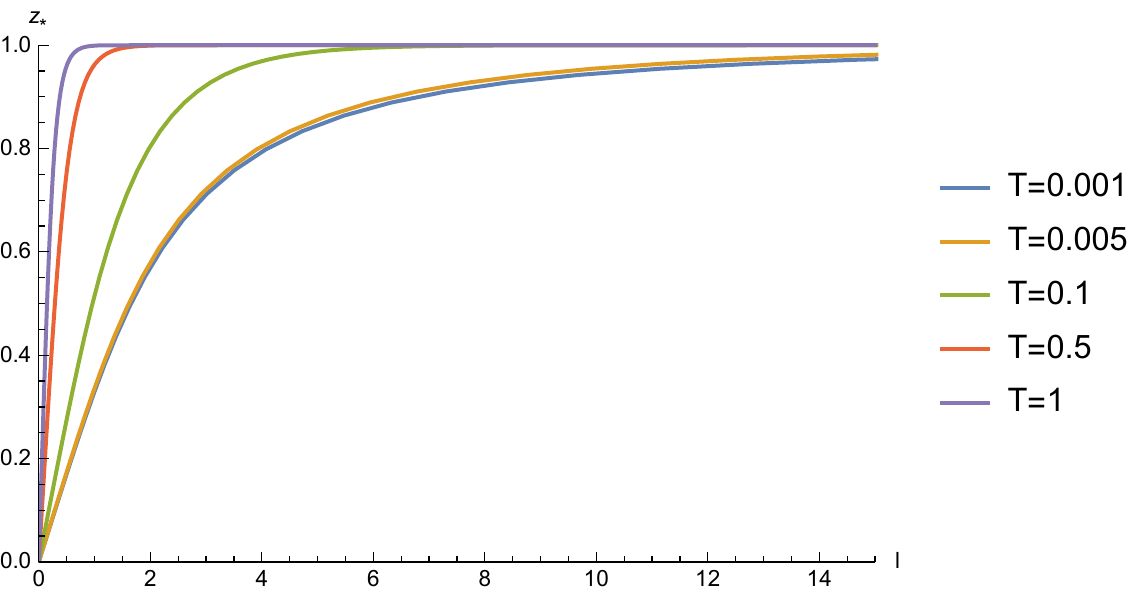}
    \includegraphics[width=0.45\textwidth]{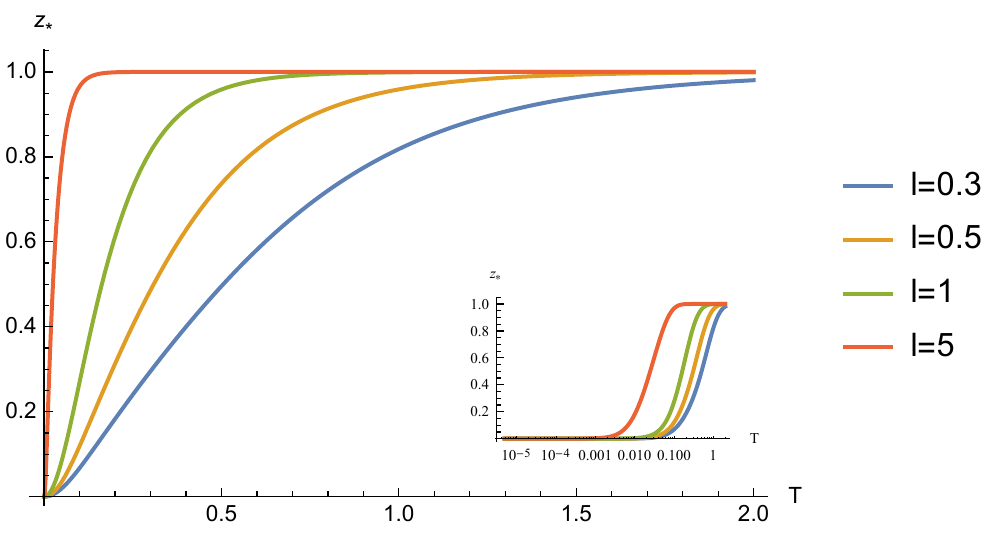}\ \hspace{0.2cm}
    \includegraphics[width=0.45\textwidth]{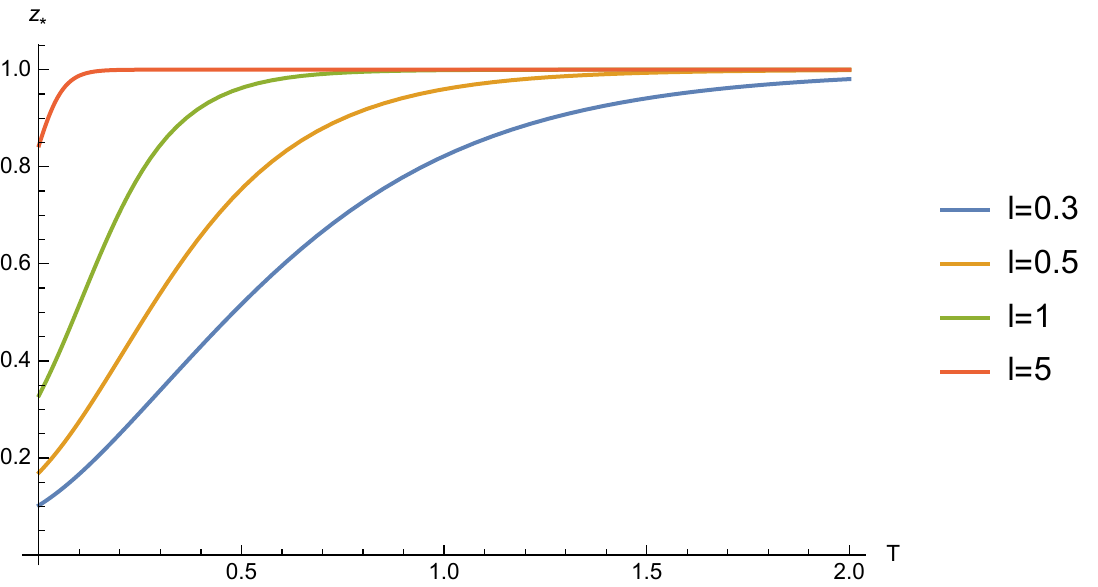}
\caption{The turning point $z_*$ as the function of width $l$ for fixed temperature (plots above) and as the function of temperature $T$ for fixed $l$ (plots below).
Left plots is the case for Gubser-Rocha model and right plots for RN-AdS background.}
    \label{fig-z_0}
\end{figure}

Before showing the behaviors of HEE, we firstly study the behaviors of the turning point $z_*$, which are exhibited in FIG.\ref{fig-z_0}.
It is easy to find that there are some obvious differences between the behaviors of $z_*$ for Gubser-Rocha model and RN-AdS geometry.
We summarize the differences and similarities between them as what follows.
\begin{itemize}
  \item In the high temperature region, $z_*$ as the function of width $l$ shares the similar behavior for Gubser-Rocha model and RN-AdS geometry.
  That is to say, $z_*$ monotonously increases as $l$ increases. When $l\rightarrow \infty$, $z_*$ approaches to the horizon of the black hole,
  while in the limit of $l\rightarrow 0$, $z_*\rightarrow 0$.
  It is expected that in high temperature limit, as we have mentioned in the above subsection,
  the $z_*$ and many information-related quantities of the Gubser-Rocha model are similar with that of the RN-AdS model.
  \item However, in the low temperature region, $z_*$ exhibits some obvious differences between Gurbser-Rocha model and RN-AdS geometry.
  From the above left plot, we see that for Gubser-Rocha model, there is a domain of $l$, where $z_*$ is almost zero.
  As $l$ increases and is beyond some critical value, $z_*$ gradually climbs up and finally approaches the black hole horizon.
  But for the RN-AdS geometry, there is no such a domain of $l$ (see the above right plot).
  Correspondingly, for Gubser-Rocha model, there is a domain of $T$, where $z_*$ almost vanishes (see the lower left plot) when $l$ is smaller than some critical value.
  But conversely for RN-AdS background, $z_*$ is finite for non-zero $l$ in the limit of zero temperature (see the below right plot).
  In low temperature region of Gubser-Rocha model,
  it is seen that for a fixed width $l$ one has $\hat l = l/\mu \to 0$, which means that the minimum surface will only stay at the near boundary region. This explains why $z_*$ for Gubser-Rocha model is significantly smaller than that for RN-AdS model in low temperature region.
\end{itemize}

We would like to point out that to find the difference of $z_*$ between Gubser-Rocha model and RN-AdS background,
we have implemented a numerical computation with higher precision. This is not an easy and straightforward work because in the limit of zero temperature, $z_*$ for Gubser-Rocha model also tends to zero, such that
much higher precision and precaution are needed in the numerics.
In addition, in the following numerical calculations, the numerical precision depends crucially on the precision of the $z_*$ value.
\begin{figure}[ht!]
    \centering
    \includegraphics[width=0.4\textwidth]{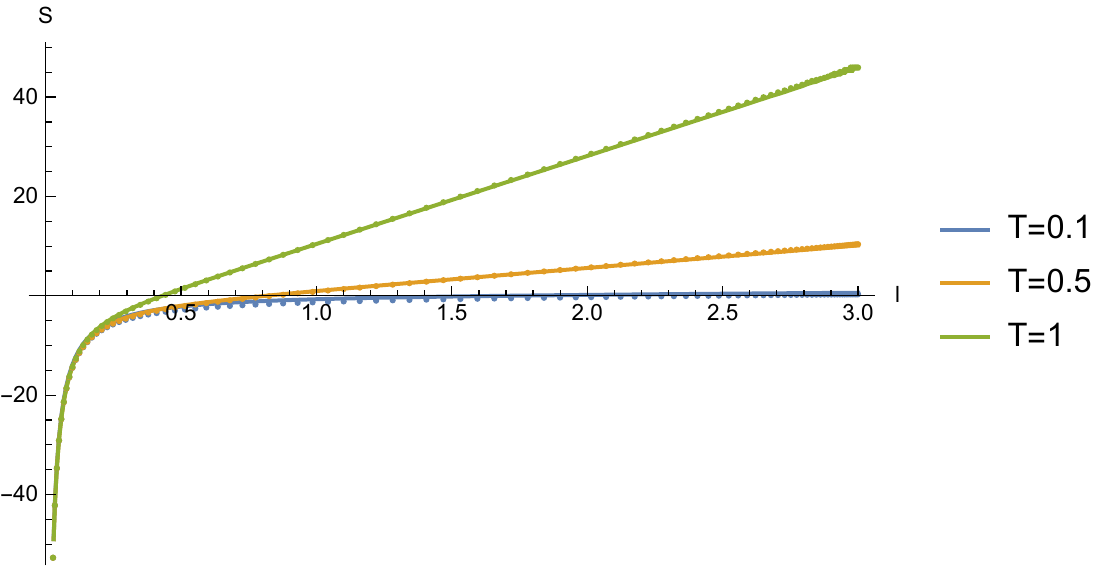}\ \hspace{2cm}
    \includegraphics[width=0.4\textwidth]{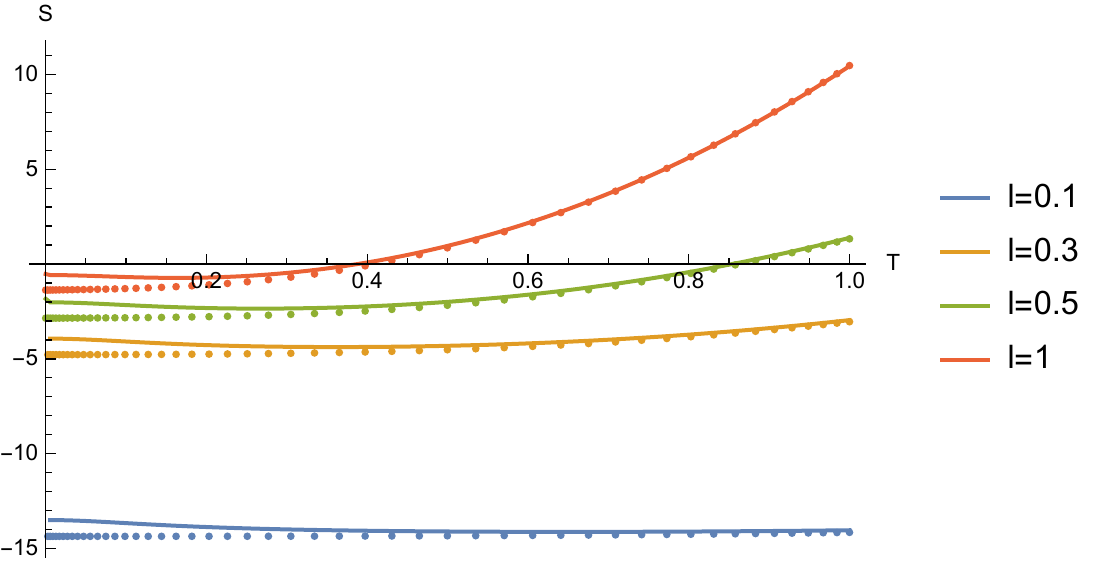}
\caption{HEE as the function of the width $l$ for fixed temperature (left plot) and as the function of temperature $T$ for fixed $l$ (right plot).
The solid lines are for the Gubser-Rocha model and the dotted lines are for the RN-AdS background.}
    \label{HEE_vs_w}
\end{figure}

We move on to study the behaviors of HEE.
The HEE as the function of width $l$ for fixed temperature and as the function of temperature $T$ for fixed $l$
are exhibited in FIG.\ref{HEE_vs_w}.
Qualitatively, the behaviors of HEE for Gubser-Rocha model are similar to that for RN-AdS background.
That is to say, for fixed and finite temperature, as the width $l$ decreases, the HEE decreases and tends to negative infinity in the limit of $l\rightarrow 0$. While for fixed width $l$, as the temperature rises, the HEE increases. However, if we take a closer look at the relation between the temperature and the HEE, we find that
in the high temperature region, the value of HEE for Gubser-Rocha model and RN-AdS background is almost the same.
But in the low temperature region, HEE for Gubser-Rocha model is larger than that for RN-AdS background.
It means that, the degree of freedom in Gubser-Rocha model is more entangled than that in the RN-AdS model in low temperature region.

\begin{figure}[ht!]
    \centering
    \includegraphics[width=0.45\textwidth]{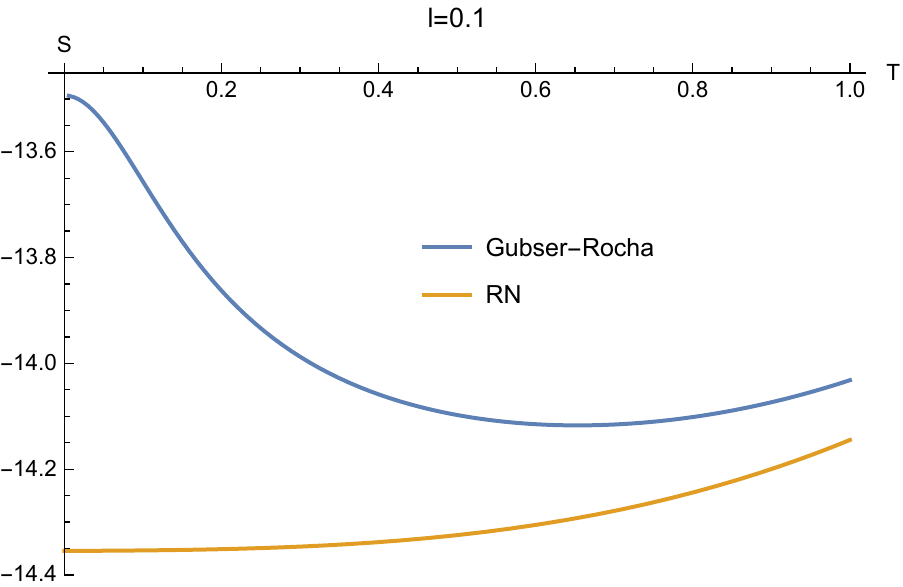}\ \hspace{0.2cm}
    \includegraphics[width=0.45\textwidth]{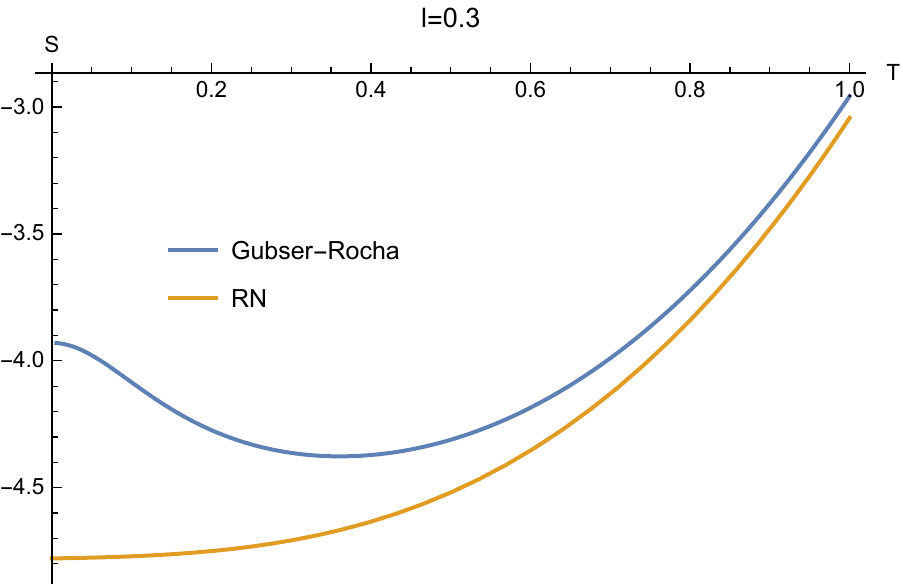}
    \includegraphics[width=0.45\textwidth]{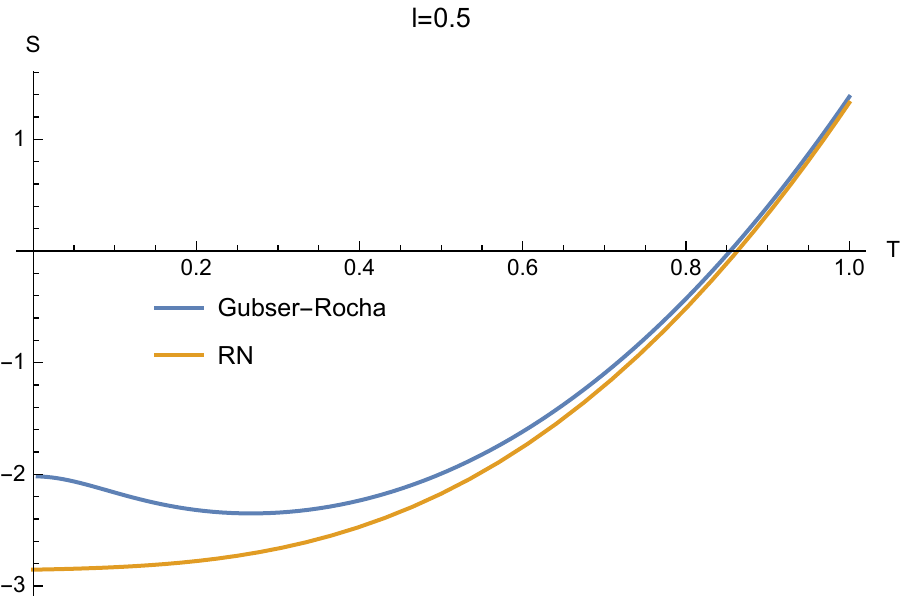}\ \hspace{0.2cm}
    \includegraphics[width=0.45\textwidth]{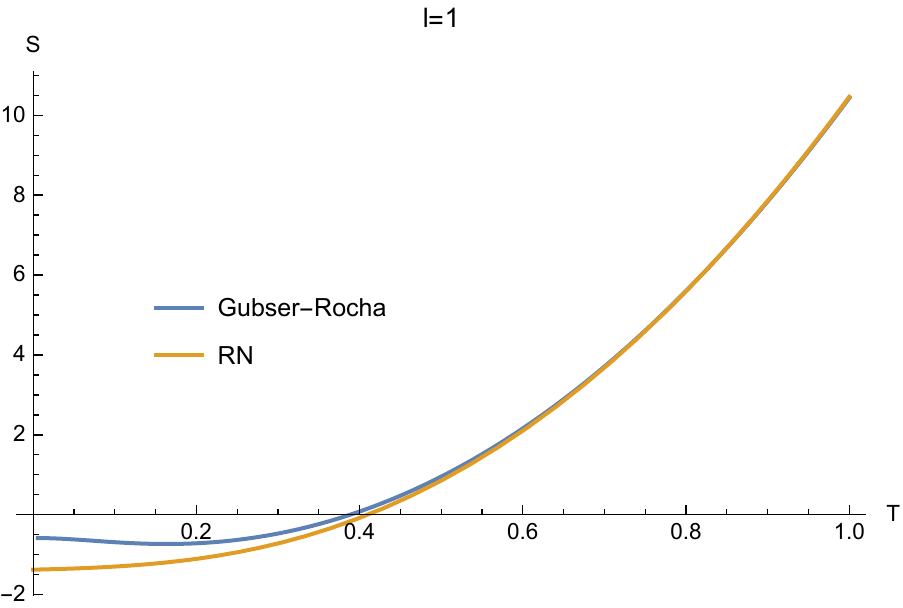}
\caption{HEE as the function of the temperature for different width $l$.
The blue line is for Gubser-Rocha model and the orange line for RN-AdS background.}
    \label{HEE_vs_tem}
\end{figure}

Further, in more details, we separately compare the Gubser-Rocha model and the RN-AdS model with different widths as FIG.\ref{HEE_vs_tem}. We find that the HEE of Gubser-Rocha model in low temperature region exhibits a non-monotonic behavior that the HEE decreases first and then increase with the increase of the temperature. But the HEE of RN-AdS model monotonically grows with the temperature, which has been analytically verified in the low temperature region in \cite{Liu:2019qje}. Here, we shall also implement an analytical understanding on the HEE behavior as the function of temperature.

Note that the HEE \eqref{S-ori} and the width \eqref{zs_vs_w} are both dimensionfull. They are also expressed in terms of the dimensionfull coordinate and metric exponents. In order to implement the temperature derivation at fixed dimensionless width $l$, we could use dimensionless $x$ coordinate at first. By the coordinate transformation $x\to \mu x$,
the dimensionless HEE can be re-casted into,
\begin{equation}\label{eq:dsdtdtfixedw}
  S = \frac{1}{\mu} \int_{0}^{\hat l} \sqrt{g_{yy}\Big(g_{x x} + g_{zz} \Big(\frac{dz}{d\hat{x}}\Big)^2\Big)}d\hat x = \frac{1}{\mu} \int_{0}^{l} \sqrt{g_{yy}\Big(g_{x x} + g_{zz} \mu^2 \Big(\frac{dz}{dx}\Big)^2\Big)}dx\,.
\end{equation}
Since $S$ is a function of $l$ and $Q$, variating the HEE with respect to $T$ renders,
\begin{equation}
  \left.\frac{\partial S}{\partial T}\right|_{l} = \Big(\frac{\partial S}{\partial l}\frac{\partial l}{\partial Q}  +\frac{\partial S}{\partial Q}\Big)\Big|_{l} \frac{dQ}{dT}\,.
  \label{eq:dsdtexv1}
\end{equation}
The first term in the bracket in the above equation can be calculated as
\fa
\Big(\frac{\partial S}{\partial l}\frac{\partial l}{\partial Q} \Big)\Big |_{l}=-\frac{l(1+2Q)}{2\sqrt{3}(Q(1+Q))^{3/2}}\,.
\ffa
In the limit of zero temperature, we have $Q\rightarrow+\infty$, therefore, this term tends to vanish and become negligible. On the other hand, we can reduce from \eqref{T} as
\begin{equation}
  \frac{dQ}{dT} = -\frac{8 \pi  Q^{3/2}}{\sqrt{3}} < 0\,.
  \label{eq:dqdt}
\end{equation}
Consequently, the sign of $\left.\frac{\partial S}{\partial T}\right|_{l}$ is determined by the sign of the $\left.\frac{\partial S}{\partial Q}\right|_{l}$. Remind that in low temperature region the minimum surface locates at the near boundary region, so we only need to consider the small $z$ case. Re-casting Eq. \eqref{eq:dsdtdtfixedw} into an integral with respect to $z$ like \eqref{zs_vs_w}, and taking the derivative, we find that up to the leading order of $z$,
\begin{equation}
  \left.\frac{\partial S}{\partial Q}\right|_{l} = \int_0^{z_*} \left[\frac{(Q+1)^3 (4 Q-1) z}{4 \sqrt{3} (Q (Q+1))^{3/2}} + \mathcal O(z^2) \right]dz  + \frac{2 Q+1}{2 \sqrt{3} (Q (Q+1))^{3/2} z_*}  > 0.
  \label{eq:dsdqv_1}
\end{equation}
Then, recalling \eqref{eq:dqdt}, it is not difficult to conclude that
\begin{equation}
  \left.\frac{\partial S}{\partial T}\right|_{l} < 0\,.
  \label{eq:dsdtv_2}
\end{equation}
Therefore, we analytically proved that in low temperature region the HEE monotonically decreases with the temperature.

For AdS-RN black hole system the zero temperature limit means $\mu \to \sqrt 6$. It indicates that the first term in the bracket in Eq.\eqref{eq:dsdtexv1} cannot be negligible even in the region of low temperature. At last it leads to $\left.\frac{\partial S}{\partial T}\right|_{l} > 0$. In contrast to the case of Gubser-Rocha model, the HEE for RN-AdS background increases as the temperature goes up even in the region of low temperature. For the detailed analysis for the case of RN-AdS background, we can refer to Ref.\cite{Liu:2019qje}.

\subsection{Mutual information}

\begin{figure}[ht!]
    \centering
    \includegraphics[width=0.5\textwidth]{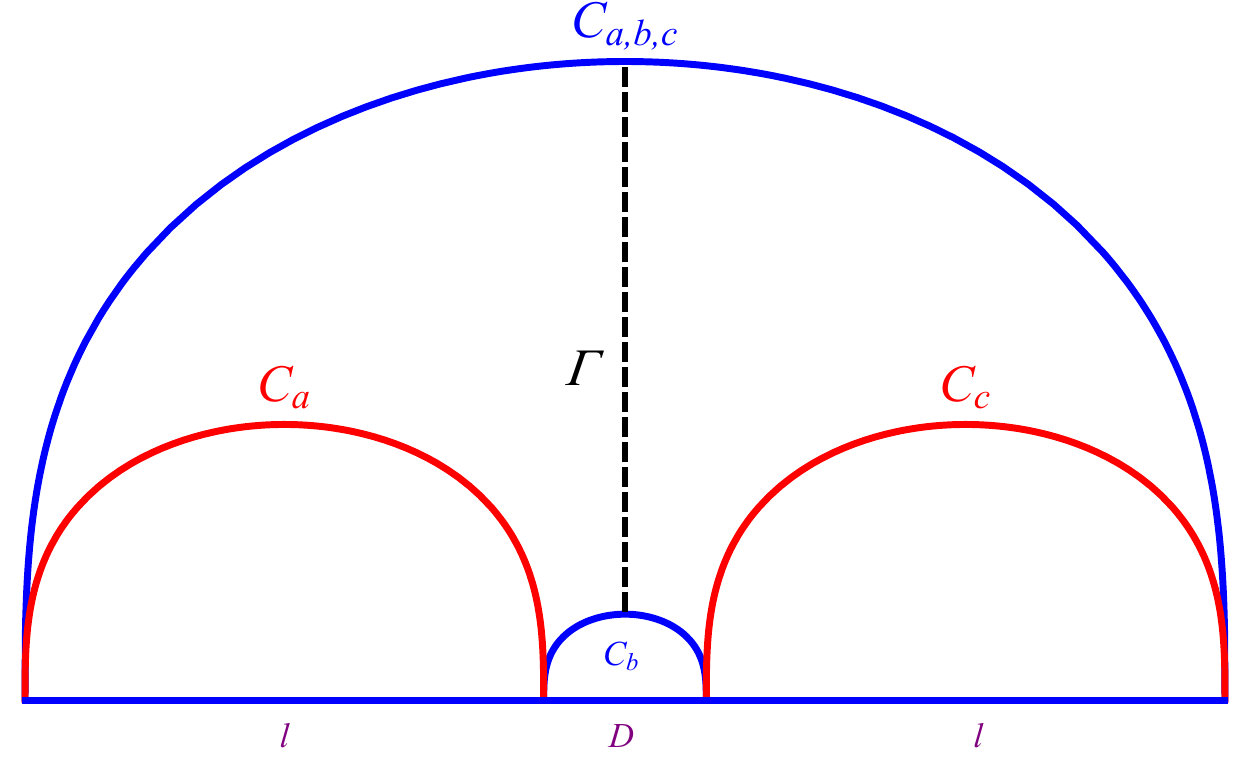}
\caption{A cartoon of MI and EoP for symmetric configuration.
A non-trivial MI equals to the difference between the area of connected configuration (blue curves)
and the area of the disconnected configuration (red curves).
The vertical dashed line $\Gamma$ represents the minimal cross section connecting the tops of $C_b$ and $C_{a,b,c}$.
$l$ denotes the size of the subsystems and $D$ is the separation scale.}
    \label{eop_cartoon_symm}
\end{figure}
In this subsection, we study the MI from Gubser-Rocha model.
HEE suffers from the divergence from asymptotic AdS boundary and so we need to introduce a cutoff, just as done as previous subsection. This issue can be avoided in MI.

To proceed, we consider two disjoint subsystems $A$ and $C$, which are separated by the subsystem $B$. Then, the MI between $A$ and $C$ can be defined as
\fa
\label{MI-def}
I(A,C)=S_A+S_C-S_{A\cup C}\,.
\ffa
A non-trivial MI requires $S(A\cup C)=S(B)+S(A\cup B\cup C)$.
Obviously, MI is a linear combination of EE.
Due to this appropriate combination, the UV divergence of HEE is removed in MI.
In addition, MI partly removes the thermal entropy contribution \cite{Fischler:2012uv}.
Therefore, it is a more relevant quantity to describe quantum entanglement.

We again consider an infinite strip configuration along $y$-axis for the subsystems and focus on the symmetric case, which is described as:
\fa
&&A:=\{0 > x > l, -\infty < y < \infty\}\,, \nonumber\\
&&B:=\{l > x > l+D, -\infty < y < \infty\}\,, \nonumber\\
&&C:=\{l+D < x < 2l+D, -\infty< y < \infty\}\,.
\ffa
$l$ is the the size of the subsystem and $D$ is separation scale.
The intersecting surface of this configuration is shown in FIG.\ref{eop_cartoon_symm}.
\begin{figure}[ht!]
    \centering
    \includegraphics[width=0.45\textwidth]{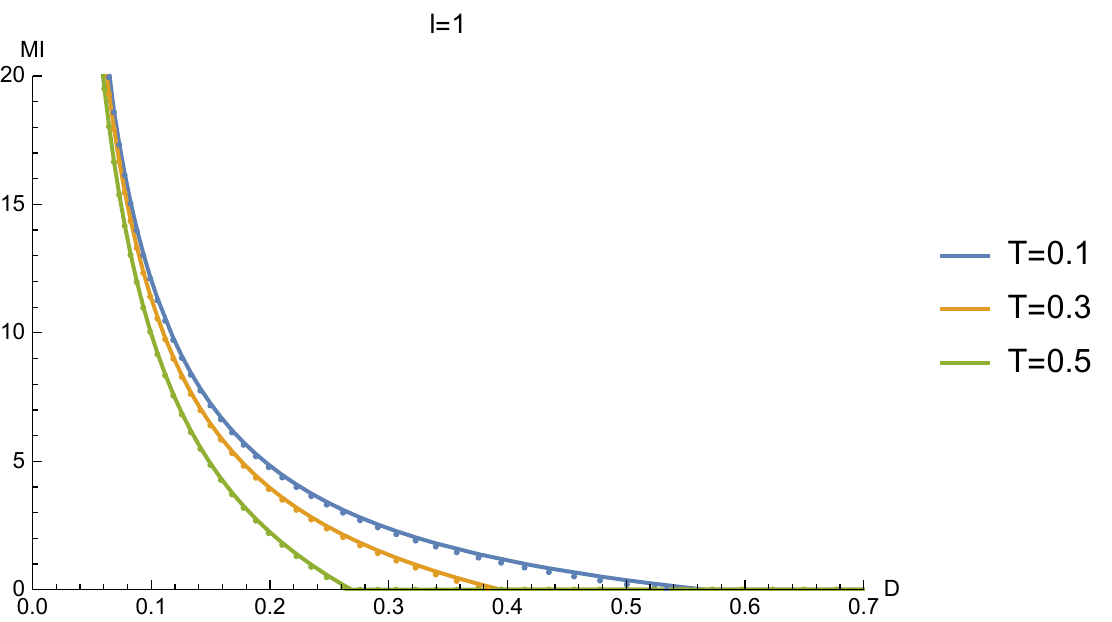}\ \hspace{1cm}
    \includegraphics[width=0.45\textwidth]{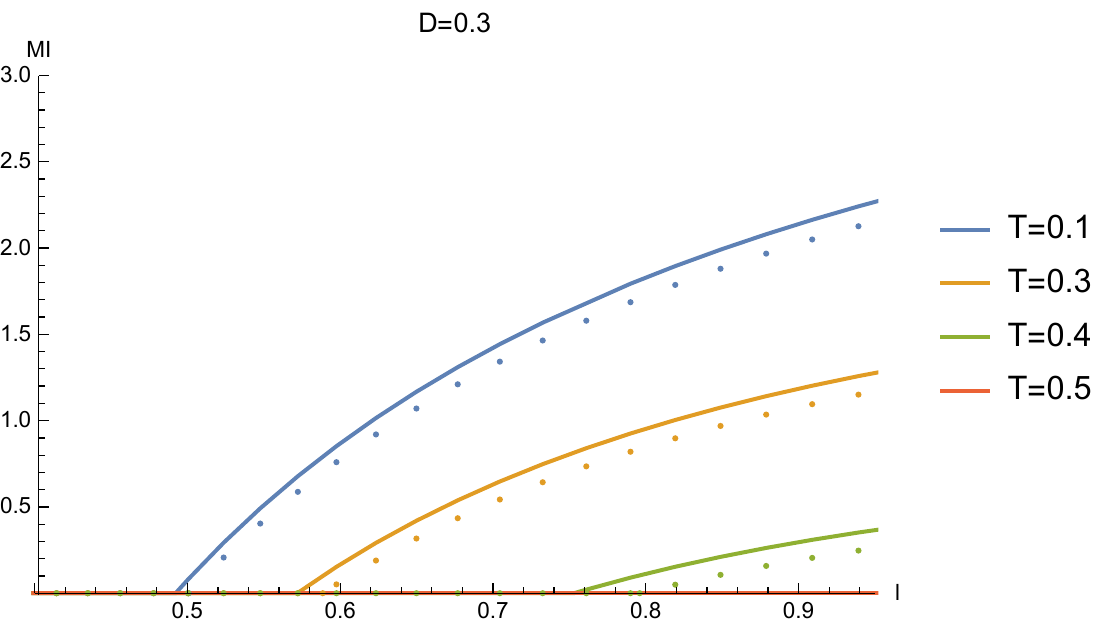}
\caption{Left plot: MI as the function of separation scale $D$ with fixed system size $l$ for different temperature.
Right plot: MI as the function of system size $l$ with fixed separation scale $D$ for different temperature.
The solid lines are for the Gubser-Rocha model and the dotted lines are for the RN-AdS background.
}
    \label{MI_sym_ab}
\end{figure}
\begin{figure}[ht!]
    \centering
    \includegraphics[width=0.5\textwidth]{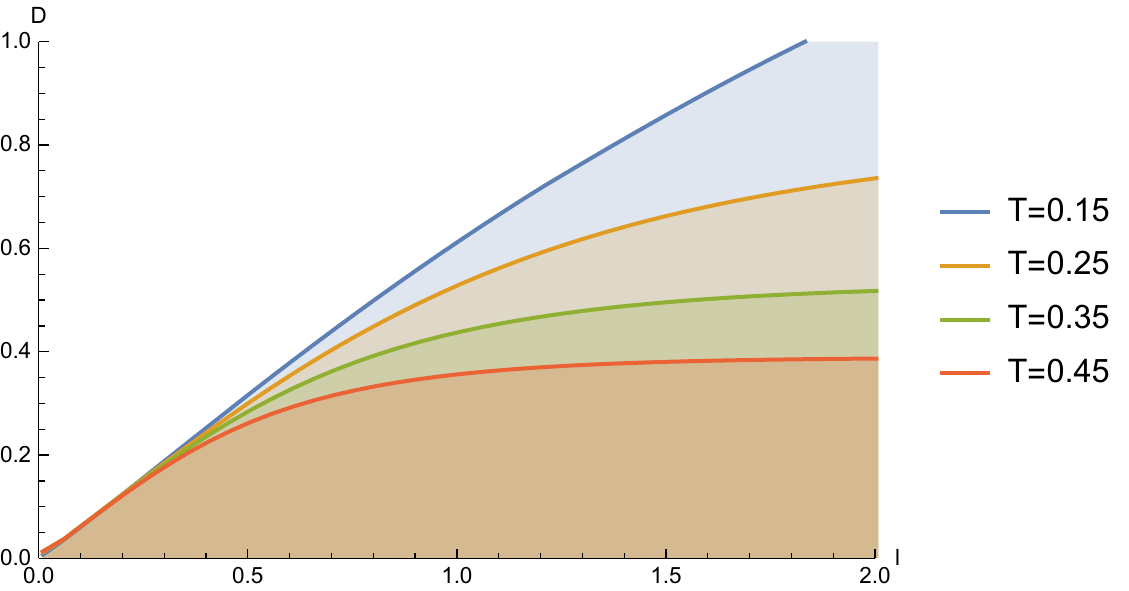}
\caption{Parameter space $(l,D)$, in which MI is non-zero only in the shaded region.
}
    \label{MI_a_b}
\end{figure}

Left plot in FIG.\ref{MI_sym_ab} exhibits MI over Gubser-Rocha model as the function of the separation scale $D$ for fixed subsystem size $l$ and different temperature (solid lines).
We find that for fixed temperature and subsystem size $l$, MI decreases as the separation scale $D$ increases.
If we further increase the separation scale such that it is beyond certain critical value, MI vanishes,
which indicates disentangling between two sub-systems.
We also exhibits the behavior of MI as the function of the system size $l$ for fixed $D$ and different temperatures in the right plot in FIG.\ref{MI_sym_ab}.
We find that MI decreases with the decrease of $l$ and vanishes when $l$ is below some value.
This result is qualitatively in agreement with that over a neutral black hole \cite{Fischler:2012uv} and also over RN-AdS background (dotted lines).
But note that quantitatively, the value of MI over RN-AdS is smaller than that over Gubser-Rocha model.
Further, FIG.\ref{MI_a_b} shows the parameter space $(l,D)$, in which the shaded region denotes non-zero MI.
An obvious characteristic is that for the fixed temperature, when the subsystem size $l$ increases, the critical lines tend to be a constant. It indicates that if we want to have a non-zero MI, the separation scale $D$ shall be constrained in certain region.

In addition, we also note that the MI of Gubser-Rocha model is always larger than that of RN-AdS model. This observation is consistent with that of the HEE, which have been found in the above section. This consistence is reasonable because the MI is directly related to the HEE.
It would be interesting to test whether another entanglement measure, the EoP, would give the same behavior. We shall discuss this question in the next subsection.

\begin{figure}[ht!]
    \centering
    \includegraphics[width=0.45\textwidth]{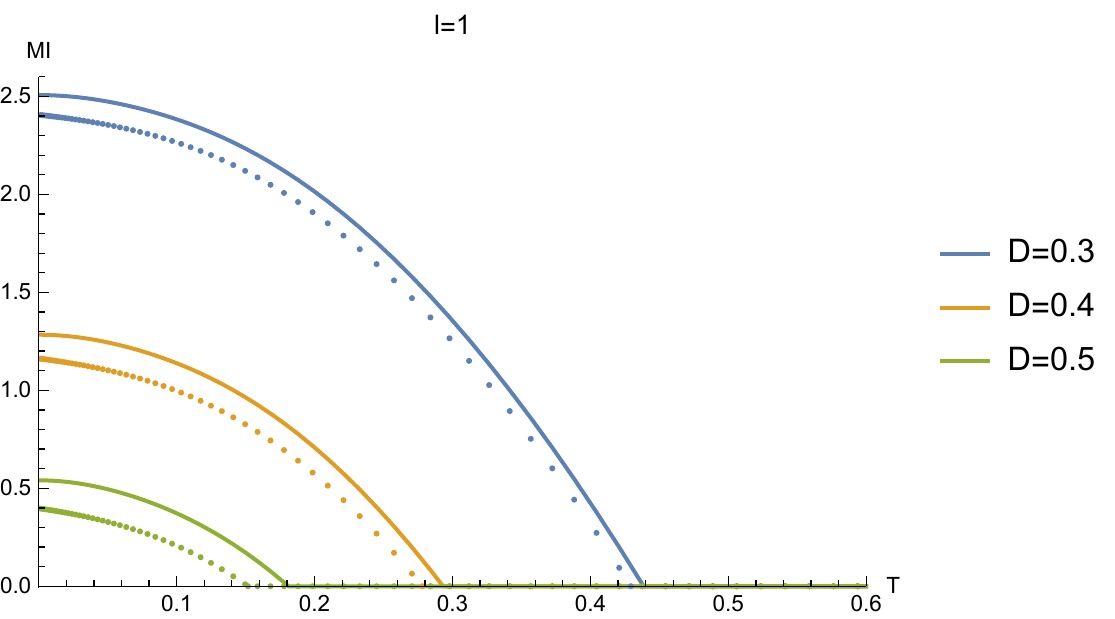}\ \hspace{1cm}
    \includegraphics[width=0.45\textwidth]{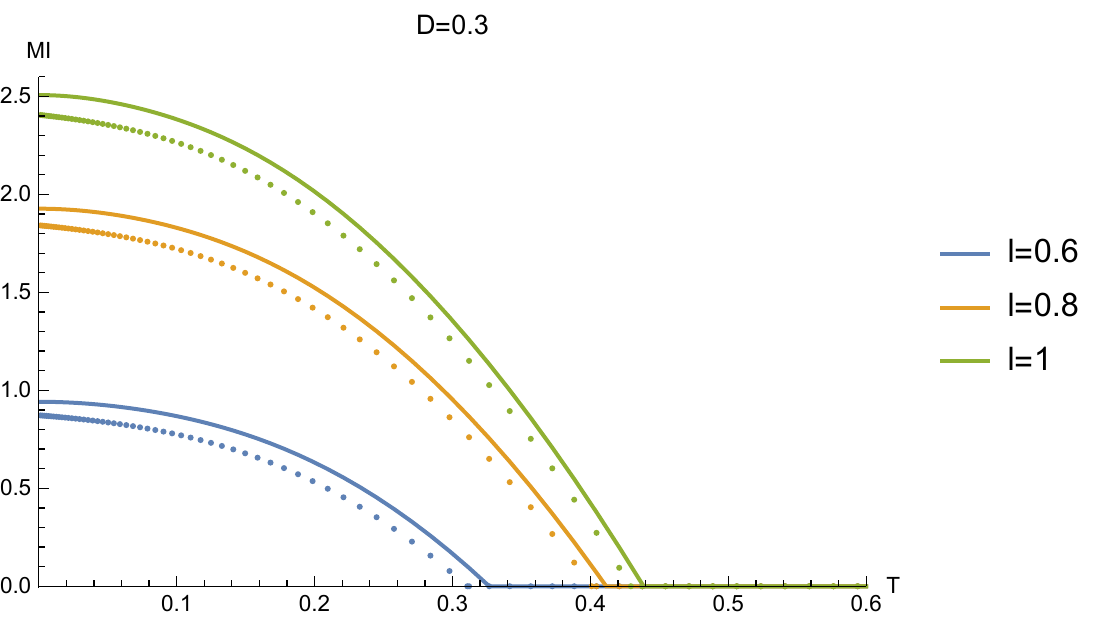}
\caption{MI as the function of the temperature for different $l$ and $D$.
The solid lines are for the Gubser-Rocha model and the dotted lines are for the RN-AdS background.}
    \label{MI_sym_tem}
\end{figure}
In the left plot in FIG.\ref{MI_sym_tem}, we show the result of how MI depends on the temperature for Gubser-Rocha model (solid lines).
We see that as the temperature rises, MI falls and finally vanishes when the temperature is beyond some critical value.
Therefore, when we heat up the system, a disentangling transition happens.
This observation is consistent with that in \cite{Fischler:2012uv,MolinaVilaplana:2011xt}.
To make a comparison, we also show MI as the function of the temperature over RN-AdS background (dotted lines in FIG.\ref{MI_sym_tem}).
We again confirm that the value of MI over RN-AdS background is smaller than that over Gubser-Rocha model.
\begin{figure}[ht!]
    \centering
    \includegraphics[width=0.5\textwidth]{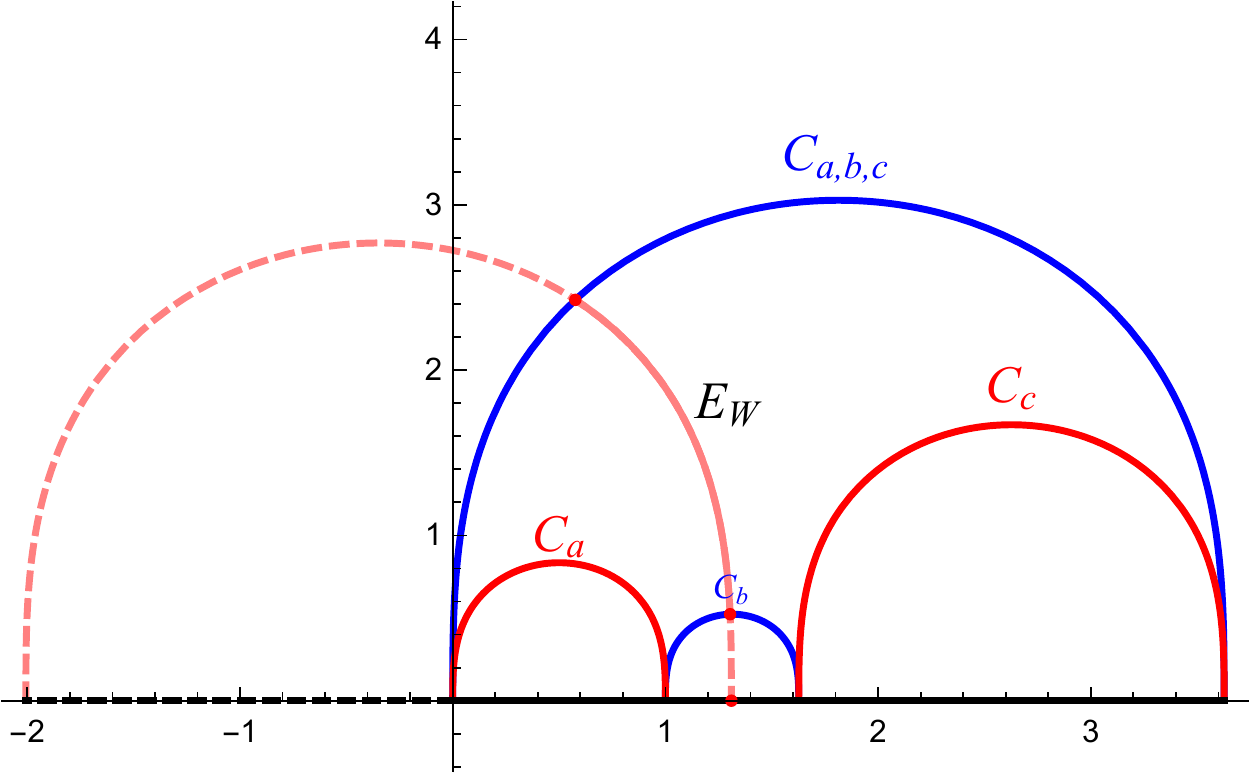}
\caption{The cartoon of MI and EoP for non-symmetric configuration.}
    \label{eop_cartoon}
\end{figure}
\begin{figure}[ht!]
    \centering
    \includegraphics[width=0.45\textwidth]{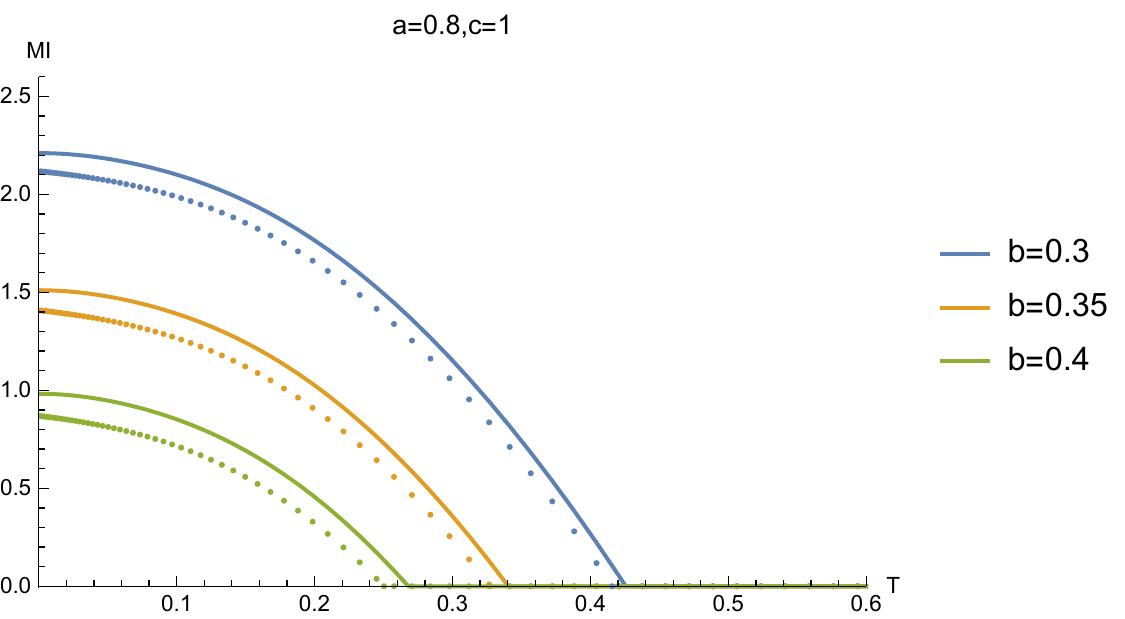}\ \hspace{1cm}
    \includegraphics[width=0.45\textwidth]{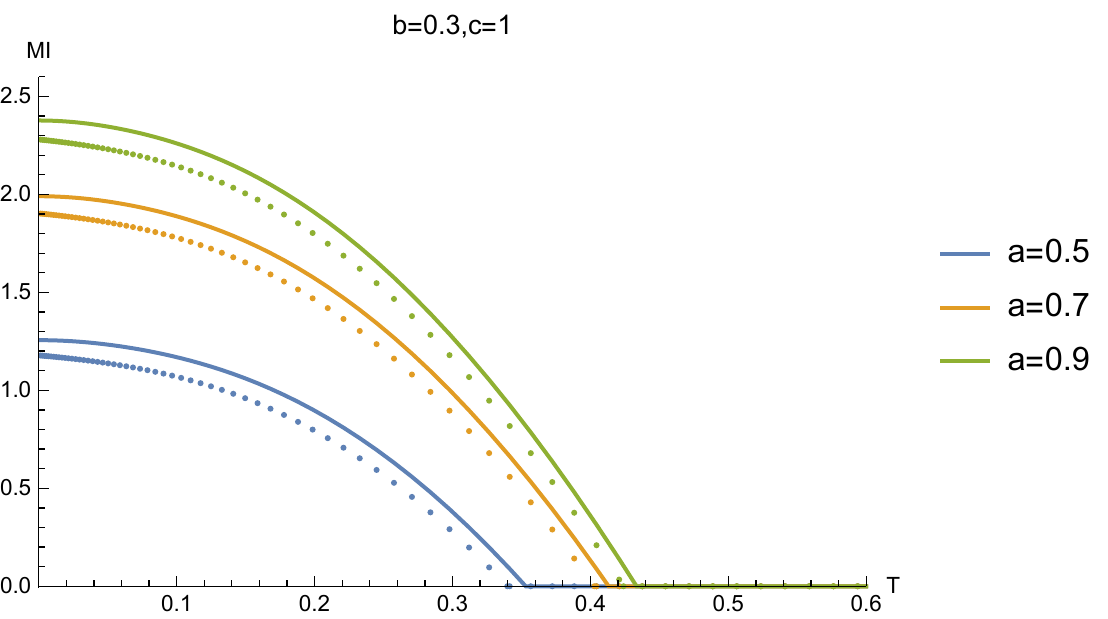}
\caption{MI as the function of the temperature for non-symmetrical configuration.
The solid lines are for the Gubser-Rocha model and the dotted lines are for the RN-AdS background.
}
\label{MI_bc}
\end{figure}

In the above study, we consider that the configuration is symmetric, i.e., the sizes of the subsystems $A$ and $C$ are equal.
Next, we turn to explore the properties of MI with non-symmetrical configuration, i.e., the sizes of $A$ and $C$ are unequal, as shown in FIG.\ref{eop_cartoon}.
We denote the sizes of the subsystems $A$ and $C$ as $a$ and $c$ and the separation size as $b$.
FIG.\ref{MI_bc} exhibits MI as the function of the temperature over Gubser-Rocha model for non-symmetrical configuration.
Qualitatively, the picture of MI for non-symmetrical configuration is consistent with that for symmetrical configuration.
In particular, the value of MI over RN-AdS geometry is also smaller than that over Gubser-Rocha model,
which indicates that this behavior is robust and independent of the configuration.

\subsection{Entanglement of purification}

In this subsection, we turn to explore EoP over Gubser-Rocha model.
EoP is an effective description on mixed state of two subsystems.
In holography, EoP is proposed as the area of the minimum cross-section of the entanglement wedge \cite{Takayanagi:2017knl,Nguyen:2017yqw}.
We firstly study the case of symmetric configuration, for which the EoP equals the area of the vertical line $\Gamma$ connecting the tops of the minimum surfaces (see FIG.\ref{eop_cartoon_symm}). The EoP over Gubser-Rocha model can be specifically calculated as
\fa \label{gamma_sym}
\Gamma&=&\int^{z_{2l+D}}_{z_D} \sqrt{g_{yy} g_{zz}} dz
\nonumber
\\
&=&\int^{z_{2l+D}}_{z_D} \frac{(16 \pi^2 \hat{T}^2 + 3z)^{3/2}}{4\pi\hat{T}z^2\sqrt{27(z-1)z^2+144 \pi^2 \hat{T}^2 z (z^2-1)+256 \pi^4\hat{T}^4(z^3-1)}} dz
\,.
\ffa
We then numerically integrate the above formula and study various properties of the EoP $E_{\omega}$.

\begin{figure}[ht!]
    \centering
    \includegraphics[width=0.45\textwidth]{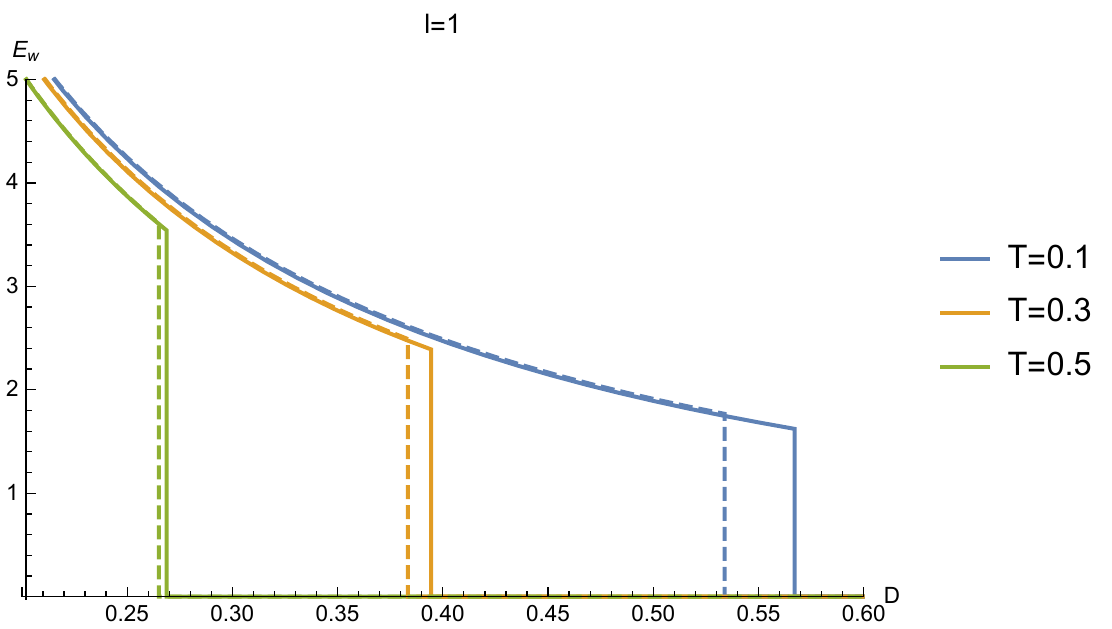}\ \hspace{1cm}
    \includegraphics[width=0.45\textwidth]{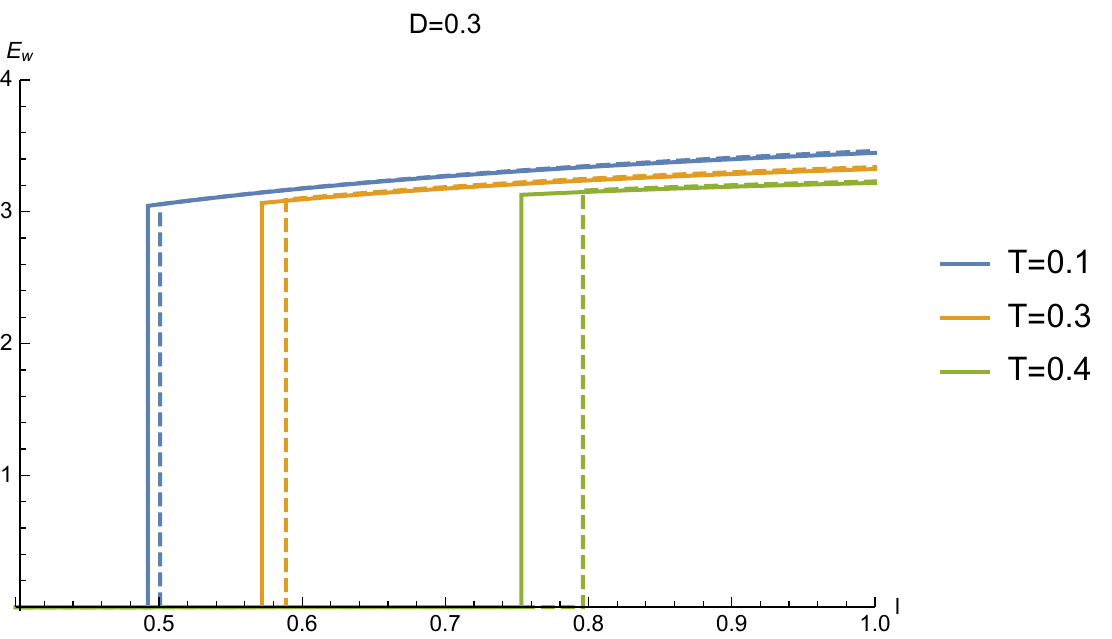}
\caption{Left plot: EoP as the function of separation scale $D$ with fixed system size $l=1$ for different temperature.
Right plot: EoP as the function of system size $l$ with fixed separation scale $D=0.3$ for different temperature temperature.
The solid lines are for the Gubser-Rocha model and the dashed lines are for the RN-AdS background.}
    \label{EoP_vs_abc_symm}
\end{figure}
Left plot in FIG.\ref{EoP_vs_abc_symm} exhibits the EoP as the function of separation scale $D$ with fixed $l=1$ for different temperatures.
It is obvious that the EoP decreases with the increase of $D$ at first,
and then, when $D$ is beyond certain critical value, EoP vanishes. It is expected  because MI vanishes and the sub-systems disentangle.
We also study the EoP as function of $l$ with fixed $D=0.3$ in the right plot in FIG.\ref{EoP_vs_abc_symm},
which shows that the EoP decreases with the decrease of $l$ and vanishes when $l$ is below some value.
At the same time, we show EoP over RN-AdS background as the function of separation scale $D$ with fixed $l=1$ for different temperatures (dashed lines).
We find that the EoP, in contrast to the MI behavior, the EoP of Gubser-Rocha model is always smaller than that of the RN-AdS model. This shows that the MI reveals the opposite entanglement property from that of the EoP. For subsystems with the same temperature and configuration, EoP shows that the dual quantum system of RN-AdS entangles strongly than that of the Gubser-Rocha model, while MI gives completely opposite conclusion. We hope that we can give a well understanding on this difference from the two informational quantities between the Gubser-Rocha model and the RN-AdS background in future.

\begin{figure}[ht!]
    \centering
    \includegraphics[width=0.45\textwidth]{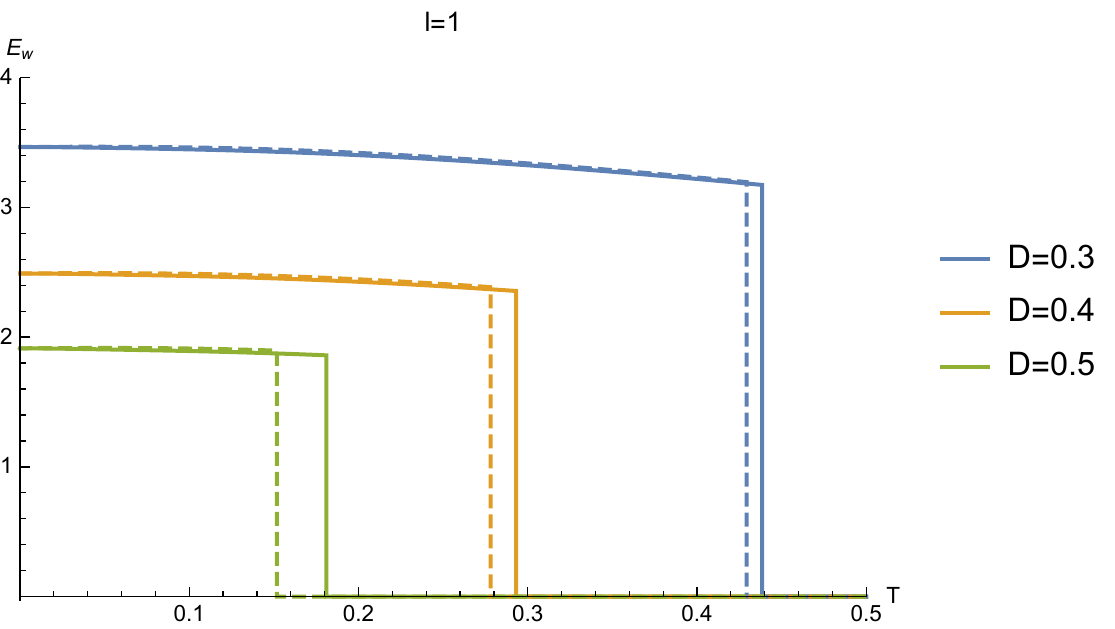}\ \hspace{1cm}
    \includegraphics[width=0.45\textwidth]{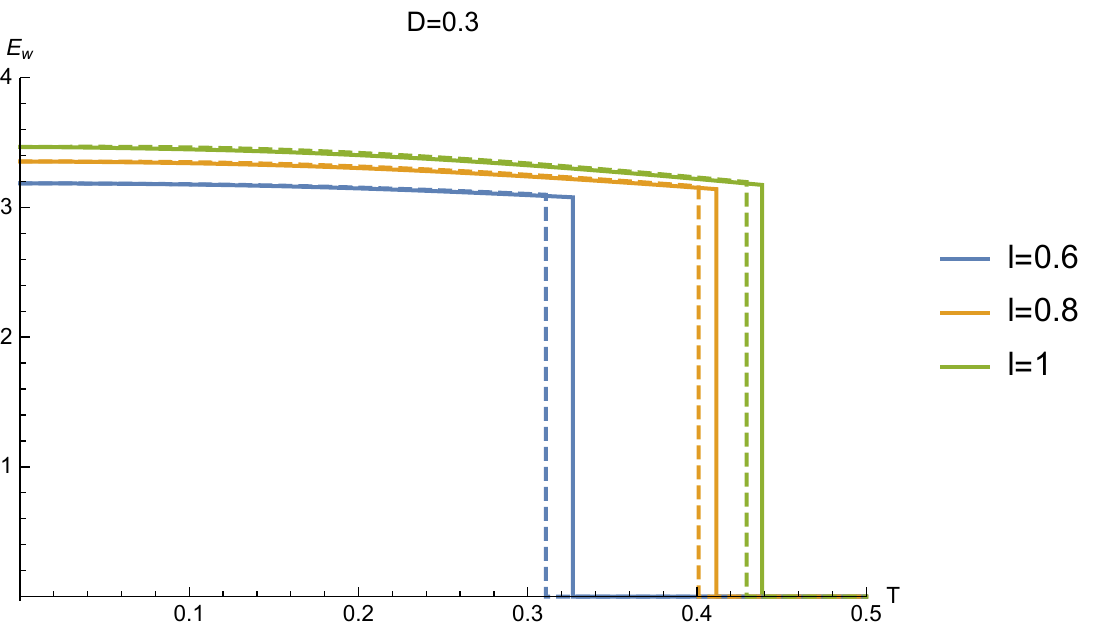}
\caption{EoP as the function of temperature $T$ with symmetric configuration.
Left plot is for different separation size $D$ and the fixed system size $l=1$.
Right plot is for the different system size $l$ and the fixed separation scale $D=0.3$.
The solid lines are for the Gubser-Rocha model and the dashed lines are for the RN-AdS background.
}
    \label{EoP_vs_Tem_symm}
\end{figure}

The temperature dependence of EoP over Gubser-Rocha model is also explored in FIG.\ref{EoP_vs_Tem_symm}.
We find that as the temperature rises, the EoP slowly decreases,
and then, when the temperature is beyond some critical value, the EoP suddenly falls to zero.
It is because the corresponding MI vanishes and means that both sub-systems is disentangle.
For comparison, we also show the temperature dependence of EoP over RN-AdS background, which is exhibited by dashed lines in FIG.\ref{EoP_vs_Tem_symm}.
Again, it confirms the observation that the EoP of Gubser-Rocha model is always smaller than that of the RN-AdS model.

\begin{figure}[ht!]
    \centering
    \includegraphics[width=0.4\textwidth]{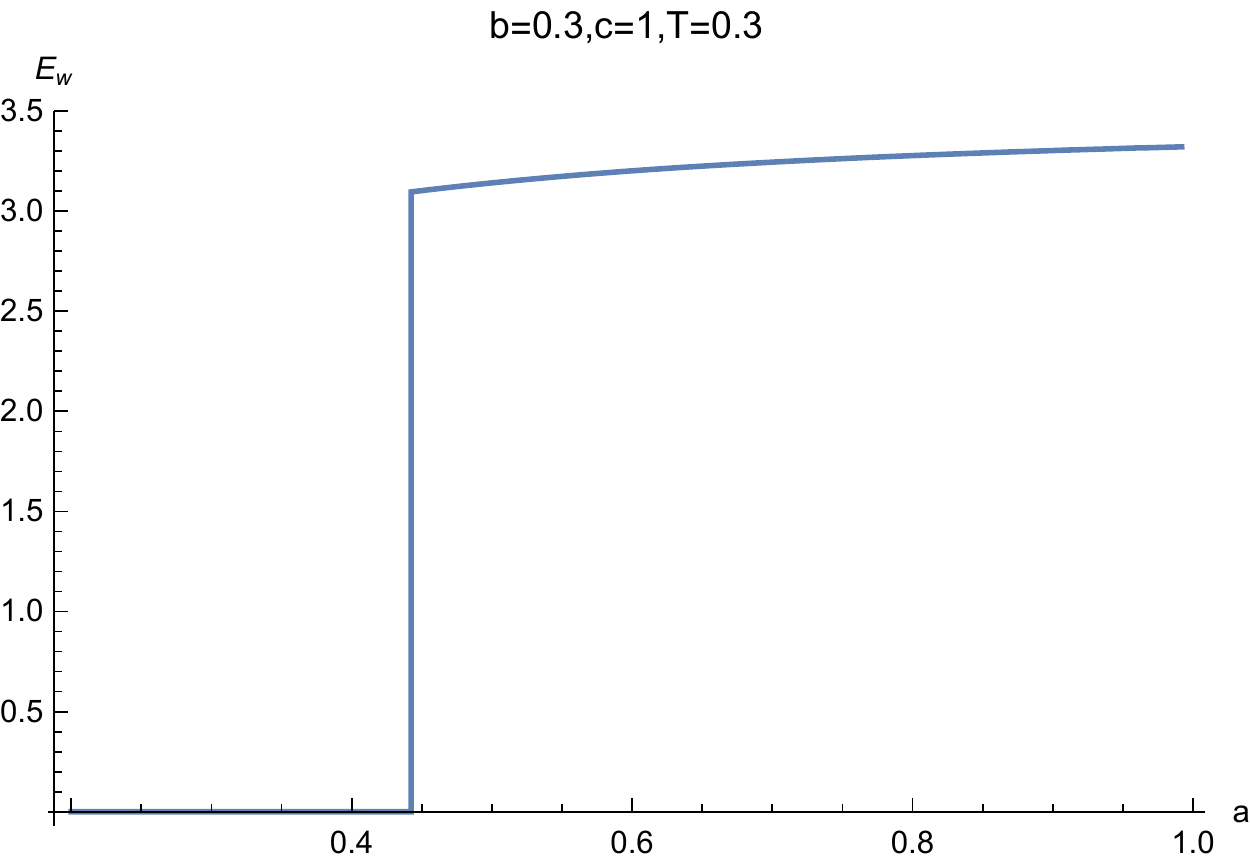}\ \hspace{1cm}
    \includegraphics[width=0.4\textwidth]{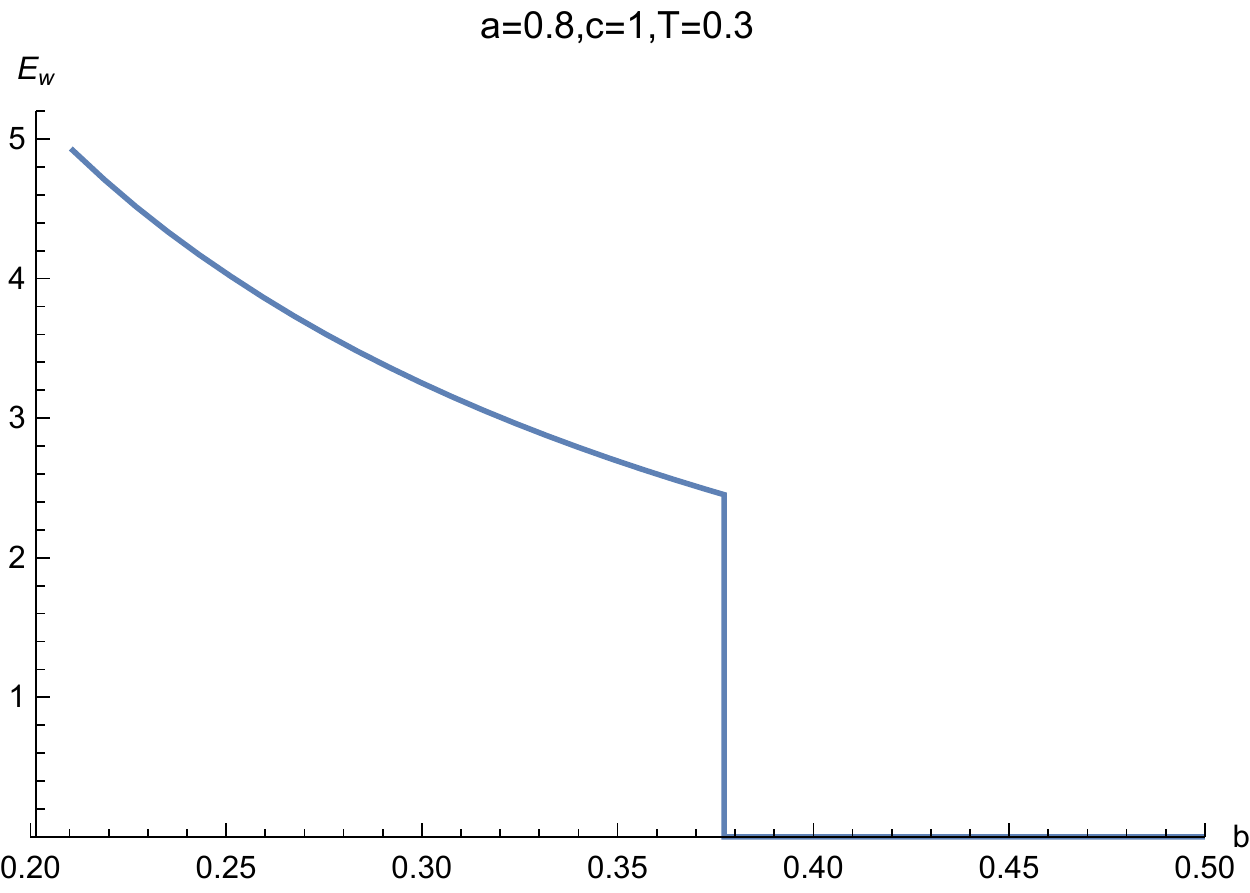}
\caption{EoP as the function of sub-system size $a$ (left plot) and the separation size $b$ (right plot), respectively.
}
    \label{eop_b&c}
\end{figure}

\begin{figure}[ht!]
    \centering
    \includegraphics[width=0.45\textwidth]{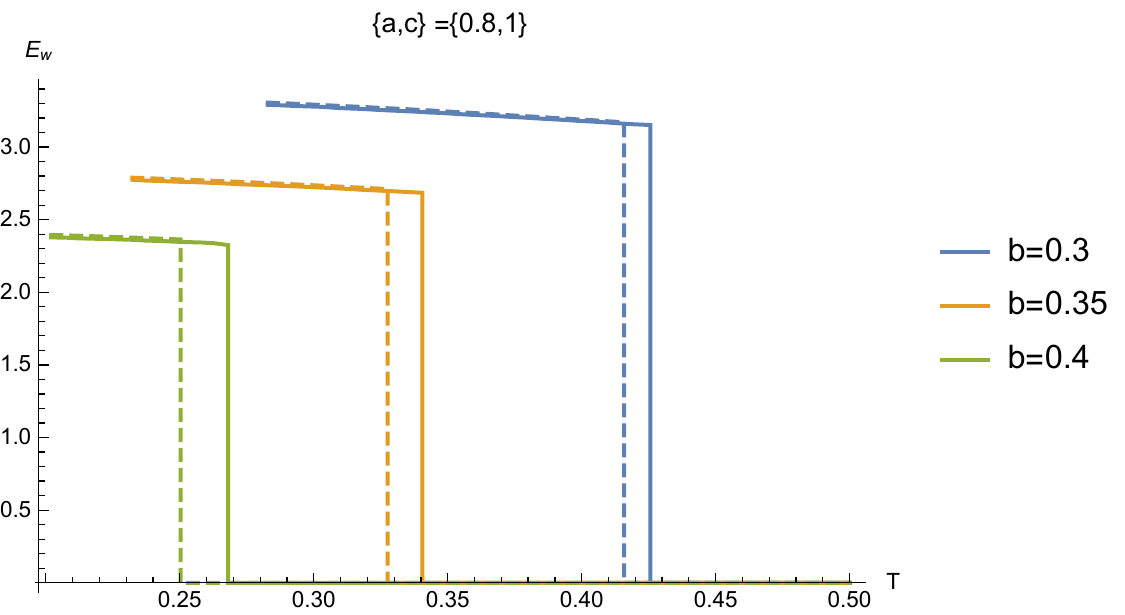}
    \includegraphics[width=0.45\textwidth]{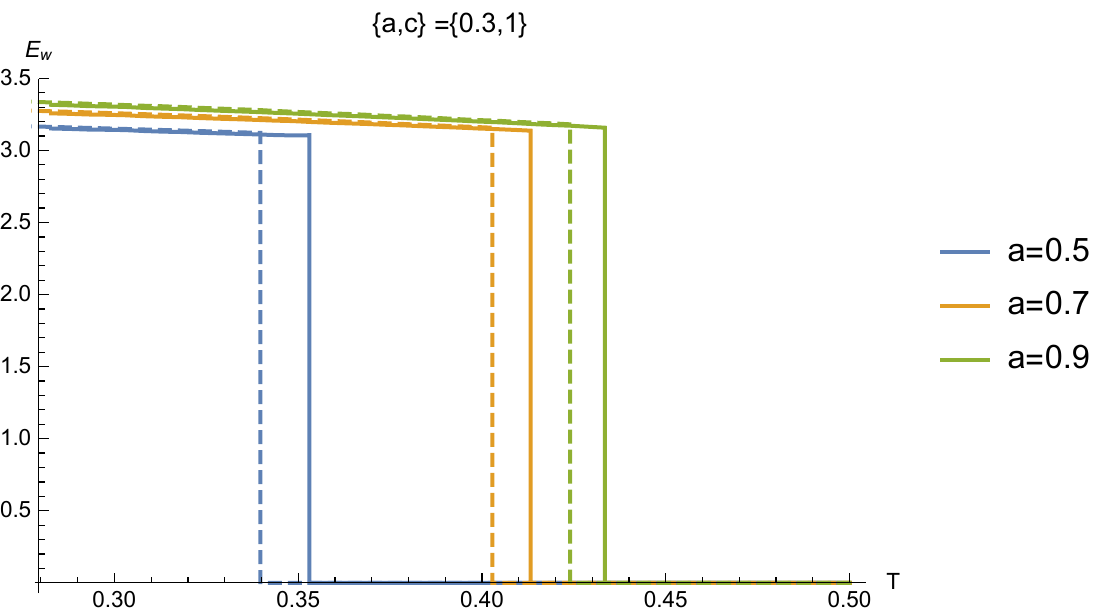}
\caption{EoP as the function of temperature for non-symmetric configuration. The solid lines are for the Gubser-Rocha model and the dotted lines are for the RN-AdS background.}
    \label{eop_tem}
\end{figure}

Next, we briefly discuss the EoP for non-symmetric configuration, for which EoP is no longer the integral between the two turning points of the sub-systems (see FIG.\ref{eop_cartoon}).
The calculation of the EoP for non-symmetric configuration is a hard work, in particular for low temperature.
Ref.\cite{Liu:2019qje} provides a detailed description on the numerical technics.
We shall follow the method provided in Ref.\cite{Liu:2019qje} to work out the EoP for non-symmetric configuration.
We sum up the results as what follows.
\begin{itemize}
  \item When the sub-system size becomes small or the separation size becomes large, both the sub-systems disentangle (FIG.\ref{eop_b&c}).
  In the entangling region, EoP monotonically increases (decreases) as the sub-system size (separation size) increases (FIG.\ref{eop_b&c}).
  \item In the high temperature region, the sub-systems are disentangling. In the intermediate temperature region,
  as the temperature rises, the EoP monotonically decreases (FIG.\ref{eop_tem}).
  These observations are consistent with that in other models, for example, RN-AdS background \cite{Liu:2019qje}
  and the holographic model with momentum relaxation \cite{Huang:2019zph,Ghodrati:2019hnn}.
  \item The EoP of Gubser-Rocha model is also smaller than that of the RN-AdS model (see FIG.\ref{eop_tem}).
\end{itemize}
The above results for non-symmetric configuration are consistent with that for symmetric configuration.

\section{Conclusion and discussion}\label{sec-con}

In this paper, we study the information quantities, including HEE, MI and EoP, over Gubser-Rocha model.
The informational quantities from Gubser-Rocha model exhibits some common characteristics as that of most holographic models.
We summarize these properties as what follows.
\begin{itemize}
  \item As the sub-system size enlarges, both MI and EoP monotonically decrease, and then when the sub-system size is beyond certain critical value,
  MI and EoP drop down to zero and so the sub-systems disentangle.
  \item When the separation size is small, the sub-systems disentangle. As the separation size increases and is beyond some certain critical value,
  both MI and EoP have non-trivial value. While the separation size is further enlarged, both MI and EoP monotonically increase.
  \item In the high temperature region, both MI and EoP monotonically decrease as the temperature climbs up, and then when the temperature exceeds some critical value,
  MI and EoP drop down to zero and a disentangling phase transition happens.
\end{itemize}

In contrast to most holographic models such as RN-AdS background with non-vanishing ground state entropy density,
Gubser-Rocha model has vanishing ground state entropy density.
We expect that some novel, even singular informational properties in the limit of zero temperature emerges.
However, we have not found any singular behavior of entanglement-related physical quantities in the limit of zero temperature limit. A brief comment is presented as what follows.
In AdS-RN black hole system the zero temperature limit means $\mu \to \sqrt 6$, where the metric and the scaling unit are both finite. Therefore, the HEE in zero temperature limit will approach a fixed value.
However, the zero temperature of Gubser-Rocha model means $Q\to \infty$ and hence $\mu \to \infty$, where the metric and the scaling unit are both infinite. The HEE behavior need to be analyzed carefully.
Interestingly, the $\mu\to\infty$ indicates that any strip with finite width $l$ will have vanishing $\hat l$. It seems like that the minimum surface should reside in the near boundary region, and the behavior of HEE seems to be determined by the AdS boundary. If it is true, the HEE will behave as $S\sim -1/l$ because in the AdS case we have $\hat S\sim -1/\hat l$ and hence we obtain $S=\hat S/\mu \sim - 1/(\mu \hat l)\sim -1/l$.
However, this expectation is not true because the zero temperature limit for Gubser-Rocha model also renders the metric being divergent.

Nevertheless, we found a peculiar property of these entanglement-related physical quantities of Gubser-Rocha model that
the HEE in low temperature region decreases with the increase of temperature, which is contrary to the entanglement property of most holographic models.
This novel phenomenon has been analytically proved and we argued that it attributes to the singular property of Gubser-Rocha model in the limit of zero temperature.

We also found that the HEE, as well as the MI, of the Gubser-Rocha model is larger than that of the RN-AdS model; meanwhile, the EoP of the Gubser-Rocha model is smaller than that of the RN-AdS model. This result suggests that, the EoP exhibits distinct behaviors from the HEE as well as the MI. In the sense of comparing the entanglement between the Gubser-Rocha model and the RN-AdS model, it is worthwhile to find out which one is the better candidate for the mixed state entanglement.

Several directions deserve further study and promotion. First of all, whether MI or EoP is more suitable for describing the mixed state entanglement is worth discussing in more holographic models. Secondly, the ground state with vanishing entropy density can be constructed based on Gubser-Rochas model, so it is worthwhile to explore the mixed state entanglement of ground states in these generalized models. Thirdly, more mixed-state entanglement, such as Renyi entropy or entanglement negativity, could be considered in this model and further compared with HEE, MI and EoP. Fourthly, as a supplement and confirmation of our numerical results, we can also analytically study the related informational quantities in different regions, especially the high/low temperature limit and the limit of large/small system scale or separated scale, following the studies \cite{Kundu:2016dyk,BabaeiVelni:2019pkw,Fischler:2012ca,Fischler:2012uv}. Finally, we can also study the non-equilibrium dynamics of related informational quantities to reveal more interesting properties of our model and further examine more inequalities of MI, EoP or reflected entropy. Lots of works along this direction have been done, see \cite{BabaeiVelni:2020wfl,Zhou:2019xzc,Zhou:2019jlh} and references therein.

\begin{acknowledgments}

This work is supported by the Natural Science Foundation
of China under Grants Nos. 11775036, 11905083, 11847055, 11705161 and Fok Ying Tung Education Foundation
under Grant No. 171006. Guoyang Fu is supported by the Postgraduate Research \& Practice Innovation Program of Jiangsu Province (KYCX20\_2973). Jian-Pin Wu is also supported by Top Talent Support Program from Yangzhou University.

\end{acknowledgments}


\begin{thebibliography}{99}

\bibitem{Maldacena:1997re}
  J.~M.~Maldacena,
  ``The Large N limit of superconformal field theories and supergravity,''
  Int.\ J.\ Theor.\ Phys.\  {\bf 38}, 1113 (1999)
  [Adv.\ Theor.\ Math.\ Phys.\  {\bf 2}, 231 (1998)]
  [hep-th/9711200].
\bibitem{Gubser:1998bc}
  S.~S.~Gubser, I.~R.~Klebanov and A.~M.~Polyakov,
  ``Gauge theory correlators from noncritical string theory,''
  Phys.\ Lett.\ B {\bf 428}, 105 (1998)
  [hep-th/9802109].
\bibitem{Witten:1998qj}
  E.~Witten,
  ``Anti-de Sitter space and holography,''
  Adv.\ Theor.\ Math.\ Phys.\  {\bf 2}, 253 (1998)
  [hep-th/9802150].
\bibitem{Aharony:1999ti}
  O.~Aharony, S.~S.~Gubser, J.~M.~Maldacena, H.~Ooguri and Y.~Oz,
  ``Large N field theories, string theory and gravity,''
  Phys.\ Rept.\  {\bf 323}, 183 (2000)
  [hep-th/9905111].

\bibitem{Maldacena:2001kr}
  J.~M.~Maldacena,
  ``Eternal black holes in anti-de Sitter,''
  JHEP {\bf 0304}, 021 (2003)
  [hep-th/0106112].
\bibitem{VanRaamsdonk:2010pw}
  M.~Van Raamsdonk,
  ``Building up spacetime with quantum entanglement,''
  Gen.\ Rel.\ Grav.\  {\bf 42}, 2323 (2010)
  [Int.\ J.\ Mod.\ Phys.\ D {\bf 19}, 2429 (2010)]
  [arXiv:1005.3035 [hep-th]].
\bibitem{VanRaamsdonk:2018zws}
  M.~Van Raamsdonk,
  ``Building up spacetime with quantum entanglement II: It from BC-bit,''
  arXiv:1809.01197 [hep-th].
\bibitem{Maldacena:2013xja}
  J.~Maldacena and L.~Susskind,
  ``Cool horizons for entangled black holes,''
  Fortsch.\ Phys.\  {\bf 61}, 781 (2013)
  [arXiv:1306.0533 [hep-th]].
\bibitem{Takayanagi:2018pml}
  T.~Takayanagi,
  ``Holographic Spacetimes as Quantum Circuits of Path-Integrations,''
  JHEP {\bf 1812}, 048 (2018)
  [arXiv:1808.09072 [hep-th]].

\bibitem{Ryu:2006bv}
  S.~Ryu and T.~Takayanagi,
  ``Holographic derivation of entanglement entropy from AdS/CFT,''
  Phys.\ Rev.\ Lett.\  {\bf 96}, 181602 (2006)
  [hep-th/0603001].
\bibitem{Takayanagi:2012kg}
  T.~Takayanagi,
  ``Entanglement Entropy from a Holographic Viewpoint,''
  Class.\ Quant.\ Grav.\  {\bf 29}, 153001 (2012)
  [arXiv:1204.2450 [gr-qc]].
\bibitem{Lewkowycz:2013nqa}
  A.~Lewkowycz and J.~Maldacena,
  ``Generalized gravitational entropy,''
  JHEP {\bf 1308}, 090 (2013)
  [arXiv:1304.4926 [hep-th]].

\bibitem{Hubeny:2007xt}
  V.~E.~Hubeny, M.~Rangamani and T.~Takayanagi,
  ``A Covariant holographic entanglement entropy proposal,''
  JHEP {\bf 0707}, 062 (2007)
  [arXiv:0705.0016 [hep-th]].
\bibitem{Dong:2016hjy}
  X.~Dong, A.~Lewkowycz and M.~Rangamani,
  ``Deriving covariant holographic entanglement,''
  JHEP {\bf 1611}, 028 (2016)
  [arXiv:1607.07506 [hep-th]].

\bibitem{Headrick:2007km}
M.~Headrick and T.~Takayanagi,
``A Holographic proof of the strong subadditivity of entanglement entropy,''
Phys. Rev. D \textbf{76} (2007), 106013
[arXiv:0704.3719 [hep-th]].
\bibitem{Wall:2012uf}
A.~C.~Wall,
``Maximin Surfaces, and the Strong Subadditivity of the Covariant Holographic Entanglement Entropy,''
Class. Quant. Grav. \textbf{31} (2014) no.22, 225007
[arXiv:1211.3494 [hep-th]].
\bibitem{Ryu:2006ef}
S.~Ryu and T.~Takayanagi,
``Aspects of Holographic Entanglement Entropy,''
JHEP \textbf{08} (2006), 045
[arXiv:hep-th/0605073 [hep-th]].

\bibitem{Casini:2011kv}
H.~Casini, M.~Huerta and R.~C.~Myers,
``Towards a derivation of holographic entanglement entropy,''
JHEP \textbf{05} (2011), 036
[arXiv:1102.0440 [hep-th]].
\bibitem{Rahimi:2016bbv}
M.~Rahimi, M.~Ali-Akbari and M.~Lezgi,
``Entanglement entropy in a non-conformal background,''
Phys. Lett. B \textbf{771} (2017), 583-587
[arXiv:1610.01835 [hep-th]].
\bibitem{Lokhande:2017jik}
S.~F.~Lokhande, G.~W.~J.~Oling and J.~F.~Pedraza,
``Linear response of entanglement entropy from holography,''
JHEP \textbf{10} (2017), 104
[arXiv:1705.10324 [hep-th]].
\bibitem{Myers:2012ed}
R.~C.~Myers and A.~Singh,
``Comments on Holographic Entanglement Entropy and RG Flows,''
JHEP \textbf{04} (2012), 122
[arXiv:1202.2068 [hep-th]].

\bibitem{Ling:2015dma}
Y.~Ling, P.~Liu, C.~Niu, J.~P.~Wu and Z.~Y.~Xian,
``Holographic Entanglement Entropy Close to Quantum Phase Transitions,''
JHEP \textbf{04} (2016), 114
[arXiv:1502.03661 [hep-th]].
\bibitem{Ling:2016wyr}
Y.~Ling, P.~Liu and J.~P.~Wu,
``Characterization of Quantum Phase Transition using Holographic Entanglement Entropy,''
Phys. Rev. D \textbf{93} (2016) no.12, 126004
[arXiv:1604.04857 [hep-th]].
\bibitem{Ling:2016dck}
  Y.~Ling, P.~Liu, J.~P.~Wu and Z.~Zhou,
  ``Holographic Metal-Insulator Transition in Higher Derivative Gravity,''
  Phys.\ Lett.\ B {\bf 766}, 41 (2017)
  [arXiv:1606.07866 [hep-th]].
\bibitem{Pakman:2008ui}
  A.~Pakman and A.~Parnachev,
  ``Topological Entanglement Entropy and Holography,''
  JHEP {\bf 0807}, 097 (2008)
  [arXiv:0805.1891 [hep-th]].
\bibitem{Kuang:2014kha}
  X.~M.~Kuang, E.~Papantonopoulos and B.~Wang,
  ``Entanglement Entropy as a Probe of the Proximity Effect in Holographic Superconductors,''
  JHEP {\bf 1405}, 130 (2014)
  [arXiv:1401.5720 [hep-th]].

\bibitem{Klebanov:2007ws}
  I.~R.~Klebanov, D.~Kutasov and A.~Murugan,
  ``Entanglement as a probe of confinement,''
  Nucl.\ Phys.\ B {\bf 796}, 274 (2008)
  [arXiv:0709.2140 [hep-th]].
\bibitem{Zhang:2016rcm}
  S.~J.~Zhang,
  ``Holographic entanglement entropy close to crossover/phase transition in strongly coupled systems,''
  Nucl.\ Phys.\ B {\bf 916}, 304 (2017)
  [arXiv:1608.03072 [hep-th]].
\bibitem{Zeng:2016fsb}
  X.~X.~Zeng and L.~F.~Li,
  ``Holographic Phase Transition Probed by Nonlocal Observables,''
  Adv.\ High Energy Phys.\  {\bf 2016}, 6153435 (2016)
  [arXiv:1609.06535 [hep-th]].
\bibitem{Guo:2019vni}
H.~Guo, X.~M.~Kuang and B.~Wang,
``Holographic entanglement entropy and complexity in St��ckelberg superconductor,''
Phys. Lett. B \textbf{797} (2019), 134879
[arXiv:1902.07945 [hep-th]].

\bibitem{Dudal:2018ztm}
D.~Dudal and S.~Mahapatra,
``Interplay between the holographic QCD phase diagram and entanglement entropy,''
JHEP \textbf{07} (2018), 120
[arXiv:1805.02938 [hep-th]].

\bibitem{Mahapatra:2019uql}
S.~Mahapatra,
``Interplay between the holographic QCD phase diagram and mutual \& $n$-partite information,''
JHEP \textbf{04} (2019), 137
[arXiv:1903.05927 [hep-th]].

\bibitem{Casini:2004bw}
H.~Casini and M.~Huerta,
``A Finite entanglement entropy and the c-theorem,''
Phys. Lett. B \textbf{600} (2004), 142-150
[arXiv:hep-th/0405111 [hep-th]].
\bibitem{Wolf:2007tdq}
M.~M.~Wolf, F.~Verstraete, M.~B.~Hastings and J.~I.~Cirac,
``Area Laws in Quantum Systems: Mutual Information and Correlations,''
Phys. Rev. Lett. \textbf{100} (2008) no.7, 070502
[arXiv:0704.3906 [quant-ph]].
\bibitem{Headrick:2010zt}
M.~Headrick,
``Entanglement Renyi entropies in holographic theories,''
Phys. Rev. D \textbf{82} (2010), 126010
[arXiv:1006.0047 [hep-th]].

\bibitem{Fischler:2012uv}
  W.~Fischler, A.~Kundu and S.~Kundu,
  ``Holographic Mutual Information at Finite Temperature,''
  Phys.\ Rev.\ D {\bf 87}, no. 12, 126012 (2013)
  [arXiv:1212.4764 [hep-th]].

\bibitem{Groisman:2004}
B. Groisman, S. Popescu, and A. Winter, ``Quantum, classical, and total amount of correlations in a quantum state'', Phys. Rev. A 72 (Sep, 2005) 032317, [quant-ph/0410091].

\bibitem{Terhal:2002}
B. M. Terhal, M. Horodecki, D. W. Leung, D. P. DiVincenzo, ``The entanglement of purification
,'' J. Math. Phys. \textbf{43}, 4286--4298 (2002), [arXiv:quant-ph/0202044].

\bibitem{Caputa:2018xuf}
P.~Caputa, M.~Miyaji, T.~Takayanagi and K.~Umemoto,
``Holographic Entanglement of Purification from Conformal Field Theories,''
Phys. Rev. Lett. \textbf{122} (2019) no.11, 111601
[arXiv:1812.05268 [hep-th]].

\bibitem{Takayanagi:2017knl}
  T.~Takayanagi and K.~Umemoto,
  ``Entanglement of purification through holographic duality,''
  Nature Phys.\  {\bf 14}, no. 6, 573 (2018)
  [arXiv:1708.09393 [hep-th]].
\bibitem{Nguyen:2017yqw}
  P.~Nguyen, T.~Devakul, M.~G.~Halbasch, M.~P.~Zaletel and B.~Swingle,
  ``Entanglement of purification: from spin chains to holography,''
  JHEP {\bf 1801}, 098 (2018)
  [arXiv:1709.07424 [hep-th]].

\bibitem{Bao:2018gck}
N.~Bao and I.~F.~Halpern,
``Conditional and Multipartite Entanglements of Purification and Holography,''
Phys. Rev. D \textbf{99} (2019) no.4, 046010
[arXiv:1805.00476 [hep-th]].
\bibitem{Umemoto:2018jpc}
K.~Umemoto and Y.~Zhou,
``Entanglement of Purification for Multipartite States and its Holographic Dual,''
JHEP \textbf{10} (2018), 152
[arXiv:1805.02625 [hep-th]].

\bibitem{Yang:2018gfq}
R.~Q.~Yang, C.~Y.~Zhang and W.~M.~Li,
``Holographic entanglement of purification for thermofield double states and thermal quench,''
JHEP \textbf{01} (2019), 114
[arXiv:1810.00420 [hep-th]].


\bibitem{Liu:2019qje}
  P.~Liu, Y.~Ling, C.~Niu and J.~P.~Wu,
  ``Entanglement of Purification in Holographic Systems,''
  JHEP {\bf 1909}, 071 (2019)
  [arXiv:1902.02243 [hep-th]].
\bibitem{Huang:2019zph}
  Y.~f.~Huang, Z.~j.~Shi, C.~Niu, C.~y.~Zhang and P.~Liu,
  ``Mixed State Entanglement for Holographic Axion Model,''
  arXiv:1911.10977 [hep-th].
\bibitem{Ghodrati:2019hnn}
  M.~Ghodrati, X.~M.~Kuang, B.~Wang, C.~Y.~Zhang and Y.~T.~Zhou,
  ``The connection between holographic entanglement and complexity of purification,''
  JHEP {\bf 1909} (2019) 009
  [arXiv:1902.02475 [hep-th]].



\bibitem{Gubser:2009qt}
S. S. Gubser, F. D. Rocha, ``Peculiar properties of a charged dilatonic black hole in $AdS_5$'', Phys. Rev. D \textbf{81}, 046001 (2010), [arXiv:0911.2898 [hep-th]].
\bibitem{Wu:2011cy}
  J.~P.~Wu,
  ``Some properties of the holographic fermions in an extremal charged dilatonic black hole,''
  Phys.\ Rev.\ D {\bf 84}, 064008 (2011)
  [arXiv:1108.6134 [hep-th]].

\bibitem{Li:2011sh}
  W.~J.~Li, R.~Meyer and H.~b.~Zhang,
  ``Holographic non-relativistic fermionic fixed point by the charged dilatonic black hole,''
  JHEP {\bf 1201}, 153 (2012)
  [arXiv:1111.3783 [hep-th]].


\bibitem{Ling:2013nxa}
  Y.~Ling, C.~Niu, J.~P.~Wu and Z.~Y.~Xian,
  ``Holographic Lattice in Einstein-Maxwell-Dilaton Gravity,''
  JHEP {\bf 1311}, 006 (2013)
  [arXiv:1309.4580 [hep-th]].

\bibitem{Alishahiha:2012ad}
M.~Alishahiha, M.~R.~Mohammadi Mozaffar and A.~Mollabashi,
``Holographic Aspects of Two-charged Dilatonic Black Hole in AdS5,''
JHEP \textbf{10} (2012), 003
[arXiv:1208.2535 [hep-th]].

\bibitem{Gubser:2000mm}
  S.~S.~Gubser and I.~Mitra,
  ``The Evolution of unstable black holes in anti-de Sitter space,''
  JHEP {\bf 0108}, 018 (2001)
  [hep-th/0011127].
\bibitem{MolinaVilaplana:2011xt}
  J.~Molina-Vilaplana and P.~Sodano,
  ``Holographic View on Quantum Correlations and Mutual Information between Disjoint Blocks of a Quantum Critical System,''
  JHEP {\bf 1110}, 011 (2011)
  [arXiv:1108.1277 [quant-ph]].

\bibitem{Kundu:2016dyk}
S.~Kundu and J.~F.~Pedraza,
``Aspects of Holographic Entanglement at Finite Temperature and Chemical Potential,''
JHEP \textbf{08} (2016), 177
[arXiv:1602.07353 [hep-th]].

\bibitem{BabaeiVelni:2019pkw}
K.~Babaei Velni, M.~R.~Mohammadi Mozaffar and M.~H.~Vahidinia,
``Some Aspects of Entanglement Wedge Cross-Section,''
JHEP \textbf{05} (2019), 200
[arXiv:1903.08490 [hep-th]].

\bibitem{Fischler:2012ca}
W.~Fischler and S.~Kundu,
``Strongly Coupled Gauge Theories: High and Low Temperature Behavior of Non-local Observables,''
JHEP \textbf{05} (2013), 098
[arXiv:1212.2643 [hep-th]].

\bibitem{Fischler:2012uv}
W.~Fischler, A.~Kundu and S.~Kundu,
``Holographic Mutual Information at Finite Temperature,''
Phys. Rev. D \textbf{87} (2013) no.12, 126012
[arXiv:1212.4764 [hep-th]].


\bibitem{BabaeiVelni:2020wfl}
K.~Babaei Velni, M.~R.~Mohammadi Mozaffar and M.~H.~Vahidinia,
``Evolution of Entanglement Wedge Cross Section Following a Global Quench,''
[arXiv:2005.05673 [hep-th]].

\bibitem{Zhou:2019xzc}
Y.~T.~Zhou, X.~M.~Kuang, Y.~Z.~Li and J.~P.~Wu,
``Holographic subregion complexity under a thermal quench in an Einstein-Maxwell-axion theory with momentum relaxation,''
Phys. Rev. D \textbf{101} (2020) no.10, 106024
[arXiv:1912.03479 [hep-th]].

\bibitem{Zhou:2019jlh}
Y.~T.~Zhou, M.~Ghodrati, X.~M.~Kuang and J.~P.~Wu,
``Evolutions of entanglement and complexity after a thermal quench in massive gravity theory,''
Phys. Rev. D \textbf{100} (2019) no.6, 066003
[arXiv:1907.08453 [hep-th]].





\end{thebibliography}
\end{document}